\documentclass[%
 reprint,
 superscriptaddress,
 amsmath,amssymb,
]{revtex4-2}

\usepackage[margin=1.0in]{geometry}

\expandafter\let\csname equation*\endcsname\relax
\expandafter\let\csname endequation*\endcsname\relax
\usepackage{amsmath}
\usepackage{mathtools}
\usepackage{amssymb}
\usepackage{graphicx}
\graphicspath{ {FigFiles/} } %% tidy subdir for figure files
\usepackage[english]{babel}
\usepackage[page,toc,titletoc,title]{appendix}
\usepackage{color,verbatim}
\usepackage{psfrag}
\usepackage{subfigure}
\usepackage{natbib}		%% to make references easier
\setcitestyle{square,numbers}
\usepackage[breaklinks=true,%
            colorlinks=true,%
            linkcolor=blue,%
            urlcolor=blue,%
            citecolor=blue]{hyperref}

\usepackage[svgnames]{xcolor}
\usepackage{tikz}

\usepackage{nicefrac}

\newcommand{\ket}[1]{|#1 \rangle}

\newcommand{\ketbra}[2]{\vert #1 \rangle \! \langle #2 \vert}

\newcommand{\average}[1]{\langle #1 \rangle}
\newcommand{\sandwich}[3]{\left \langle #1 \mid #2 \mid #3 \right\rangle}

\newcommand{\dd}[2]{\ensuremath{\frac{\mathrm{d} #1 }{\mathrm{d} #2 }}}

\usepackage{bbold}   %% for a nice identity symbol

\begin{document}

\title{An energetic perspective on rapid quenches in quantum annealing}

\author{Adam Callison}
\email{a.callison16@ic.ac.uk}
\affiliation{Blackett Laboratory, Imperial College London, London SW7~2BW, UK}

\author{Max Festenstein}
\affiliation{Blackett Laboratory, Imperial College London, London SW7~2BW, UK}
\affiliation{Department of Physics; Joint Quantum Centre (JQC) Durham-Newcastle, Durham University, South Road, Durham, DH1~3LE, UK}

\author{Jie Chen}
\affiliation{Department of Physics; Joint Quantum Centre (JQC) Durham-Newcastle, Durham 
University, South Road, Durham, DH1~3LE, UK}

\author{Laurentiu Nita}
\affiliation{Department of Physics; Joint Quantum Centre (JQC) Durham-Newcastle, Durham 
University, South Road, Durham, DH1~3LE, UK}

\author{Viv Kendon}
\affiliation{Department of Physics; Joint Quantum Centre (JQC) Durham-Newcastle, Durham 
University, South Road, Durham, DH1~3LE, UK}

\author{Nicholas Chancellor}
\email{nicholas.chancellor@gmail.com}
\affiliation{Department of Physics; Joint Quantum Centre (JQC) Durham-Newcastle, Durham 
University, South Road, Durham, DH1~3LE, UK}

\date{\today}

\begin{abstract}
There are well developed theoretical tools to analyse how quantum dynamics can solve computational problems by varying Hamiltonian parameters slowly, near the adiabatic limit.  On the other hand, there are relatively few tools to understand the opposite limit of rapid quenches, as used in quantum annealing and (in the limit of infinitely rapid quenches) in quantum walks.  In this paper, we develop several tools which are applicable in the rapid quench regime.  Firstly, we analyse the energy expectation value of different elements of the Hamiltonian. From this, we show that monotonic quenches, where the strength of the problem Hamiltonian is consistently increased relative to fluctuation (driver) terms, will yield a better result on average than random guessing.  Secondly, we develop methods to determine whether dynamics will occur locally under rapid quench Hamiltonians,
and identify cases where a rapid quench will lead to a substantially improved solution.  
In particular, we find that a technique we refer to as ``pre-annealing'' can significantly improve the performance of quantum walks.
We also show how these tools can provide efficient heuristic estimates for Hamiltonian parameters, a key requirement for practical application of quantum annealing.

\end{abstract}

\maketitle %%%%%%%%%%%%%%%%%%%%%%%%%%%%%%%%%%%%%%

\tableofcontents %%%%%%%%%%%%%%%%%%%%%%%%%%%%%%%

\section{Introduction} %%%%%%%%%%%%%%%%%%%%%%%%%

Quantum computing using continuous time evolution has gained much interest in recent years.  This includes adiabatic quantum computing \citep{farhi00a}, quantum annealing \citep{finilla94a,kadowaki98a}, and continuous-time quantum walks \citep{farhi98a}.
Optimisation tasks are a natural application for quantum computing in this setting, and  have been explored in many diverse fields including traditional computer science \citep{chancellor16a,choi10b,choi10c},  decoding communications \citep{chancellor16b}, finance \citep{marzec16a,Orus18a,Venturelli18a}, error correction of quantum memories \citep{Roffe19a}, scheduling \citep{Venturelli15a,crispin13a,Tran16a}, computational biology \citep{perdomo-ortiz12a}, flight gate assignment \citep{Stollenwerk18a}, air traffic management \citep{Stollenwerk19a}, and hydrology \citep{omalley18a}. 
This is partially due to advances in the theoretical foundations of adiabatic quantum computing, including proofs that it is universal in certain settings \cite{kempe06,aharonov09}, improved versions of the adiabatic condition \citep{Amin09adiabat,Cheung11adiabat,lidar09a}, and an extension of the adiabatic theorem to open systems \citep{Venuti16open}.
For a comprehensive review of these and other advances, see Albash and Lidar \citep{albash16a}. 
More recently, it has been shown by Hastings \cite{hastings2020power} that, even when no sign problem exists, there is a superpolynomial oracle separation between adiabatic quantum computing and classical computing. Other recent advances have come from new ways to map problems, for instance the methods of encoding more connected graphs than the native hardware connections using parity, often called parity AQC \citep{Lechner15a,Rocchetto16a,Leib16a,chancellor17a}.  These provide an alternative to the more traditional minor embedding techniques \citep{choi08a,choi10a}, and may be easier to implement experimentally.

In this paper, we focus on the coherent regime of operation, for which the effects of thermal dissipation and decoherence can be neglected. Such a regime could be experimentally reached either by reducing noise, implementing quantum error correction
\citep{Jordan06a,sarovar05a,young13a,Sarovar13a,Freeman18a,Atalaya20a,Pudenz14a,Bookatz15a,Lidar08a}, or quenching on a timescale which is much faster than the decoherence time. 
A complementary approach to reduce noise is to implement dynamics which reduce or eliminate the interaction between the system and its environment through quantum interference effects, known as dynamical decoupling \citep{Lidar08a,Paz-Silva12a,Quiroz12a}. 
Although current superconducting quantum annealing hardware operates in a dissipative regime \cite{dickson13a}, quantum annealing has been implemented in atomic settings where coherence is easier to maintain than in superconducting circuits \citep{Bernien2017a}, and progress has been made to reduce noise in superconducting circuit settings \citep{low_noise_whitepaper}. 
There have been experimental implementations of simple forms of error correction in quantum annealing \citep{Pudenz14a,Pudenz15a,Vinci15a,Vinci2016a,Vinci18a}, and efforts have been made to circumvent experimental limitations on quench rates in superconducting systems \citep{lanting_AQC2018}.

In a fully coherent regime, the dynamics are straightforward to model theoretically, since they can be described by a set of qubits (two state quantum systems) under the action of a Hamiltonian, evolving according to the Schr\"odinger equation. Conventionally, the Hamiltonian for this evolution is written as the sum of a \emph{problem} Hamiltonian $H_{\mathrm{prob}}$, which is diagonal in the computational basis, and encodes the classical problem being solved, and a \emph{driver} Hamiltonian $H_{\mathrm{drive}}$ which implements quantum dynamics to explore the solution space.
We use two equivalent forms for the total Hamiltonian.  First,
\begin{equation}
H_{\mathrm{AB}}(t)=A(t)\,H_{\mathrm{drive}}+B(t)\,H_{\mathrm{prob}}, \label{eq:Hopt}
\end{equation}
where $A(t)$ and $B(t)$ are positive, time-dependent control functions.
However, typically the crucial feature is what happens to the ratio of driver to problem strength $\nicefrac{A(t)}{B(t)}$ as the algorithm progresses.
As such, we define an alternative parametrization of the Hamiltonian, up to an overall (time-dependent) scaling factor $B(t)$, as
\begin{equation}
H_{\Gamma}(t)=\Gamma(t)\,H_{\mathrm{drive}}+H_{\mathrm{prob}}, \label{eq:Hopt_Gamma}
\end{equation}
where there is a single control function $\Gamma(t)>0$ for the ratio $\nicefrac{A(t)}{B(t)}$.
Since (\ref{eq:Hopt}) and (\ref{eq:Hopt_Gamma}) are equivalent, up to a rescaling of the time parameter, results for one form of Hamiltonian will generalize to results for the other.  We use both forms, choosing the most convenient for the specific problem or example.

Hamiltonians of the form (\ref{eq:Hopt}) and (\ref{eq:Hopt_Gamma}), which begin with $A(t)>0$ and $B(t)=0$ and end with $A(t)=0$ and $B(t)>0$, or equivalently, begin with $\Gamma(t)\gg1$ and end with $\Gamma(t)=0$, are used for most types of continuous-time quantum computing.
When run on a much shorter timescale than required for adiabatic quantum computing, we call this a \emph{rapid quench}. 

The simplest form of continuous-time quantum computing in the coherent regime is \emph{continuous time quantum walk} (QW) introduced by \cite{farhi98a,childs03a}, in which the control functions are time-independent and set so that $\Gamma(t)=\gamma$ where $\gamma$ is a constant hopping rate.
This can be viewed as the limit of an infinitely fast quench, in which $B(0)$ jumps from zero to $A(0)/\gamma$ at $t=0$ and $A(t_f)$ drops to zero at the final time $t_f$.
The other pure state continuous-time quantum computing which is commonly considered is \emph{adiabatic quantum computing} (AQC) introduced by \cite{farhi00a}, for which the control functions $A(t)$ and $B(t)$ are varied slowly from $A(0)=1$ and $B(0)=0$ to $A(t_f)=0$ and $B(t_f)=1$.
By the adiabatic theorem of quantum mechanics, this achieves a success probability (probability of finding the ground state of the problem Hamiltonian $H_{\mathrm{prob}}$) which approaches $1$ as $t_f\rightarrow \infty$. 
For a review of AQC see \cite{albash16a}. For a thorough discussion of the relationship between AQC and QW, see the introductions of \cite{Morley19a,callison19a}.
The fully coherent  regime has provable quantum speedups in the case of both AQC and QW.
For instance, unstructured search, the continuous time analog of Grover's search, can yield the same speedup in the AQC \citep{roland02a} and QW \citep{childs03a} settings as the gate based counterpart. It is possible to interpolate between these two techniques while preserving the speedup \citep{Morley19a}.

For problems which are closer to real world optimisation, theoretical studies have mostly focused on AQC \citep{albash16a}, likely because the adiabatic theorem provides a general way to show that such algorithms could in principle succeed with high probability.
While theoretically tractable, the adiabatic regime is difficult to reach experimentally, and contains some counter-intuitive effects in the deep adiabatic regime \citep{weibe12a,CamposVenuti18a,Passos19a}.
Solving NP-hard problems adiabatically will at most obtain a polynomial speed up (assuming $\mathrm{P}\ne\mathrm{NP}$).
Since AQC requires the system to remain coherent throughout, exponentially long runtime requires exponentially long coherence time, which is experimentally challenging for near-term quantum computing.
When the runtime is limited by a constant or mildly scaling coherence time, such an algorithm could only solve the problem with an exponentially low probability, and therefore require exponentially many repeats to succeed with high probability. This approach, however, is a valid one for problems other than search. Recent numerical results on spin-glasses using QW show favourable scaling from many short run repeats \citep{callison19a}. It has also been numerically demonstrated that rapid quenches can be superior to long quenches for AQC-like algorithms \citep{Crosson14a}. 
Recently, Crosson and Lidar \cite{crosson2020prospects} made an important contribution to the theory of quantum annealing outside of the adiabatic limit by introducing \textit{diabatic} quantum annealing (DQA), which formalizes ideas described in \cite{muthukrishnan16,katsuda2013nonadiabatic}. DQA relies on a generalisation of the adiabatic theorem from \cite{jansen07}, and describes a class of quantum annealing algorithms in which amplitude is restricted to a low-energy part of the Hamiltonian spectrum.

Finally, for single shot, high success probability algorithms for NP-hard problems, achieving even a polynomial speedup typically requires setting, with exponential precision, the control functions to values which lead to exponential small gaps in the Hamiltonian spectrum.
This was shown to be necessary for unstructured search in \cite{childs02a,roland02a,Morley19a} and for the random energy model \citep{Farhi08a} in \cite{callison19a}.
This requirement is problematic, as there are no general methods for determining where these gaps occur and because such precise control settings can be difficult to achieve in real hardware.
Recent work by Chakraborty et al \cite{chakraborty2018finding} demonstrates that some of the fine tuning requirements in unstructured search can be avoided by formulating the Hamiltonian differently; it is unclear whether this approach would extend to the random energy model of \cite{Farhi08a}.

Given the near term importance of methods which can succeed with limited coherence time, in this paper, we develop mathematical tools to increase our understanding of how computation is achieved in both the rapid quench regime and quantum walks.
These tools are important not only for theoretical understanding of when adiabatic algorithms and rapid quenches will be effective, but also for choosing parameters for the Hamiltonians used. 
While some theoretical arguments \citep{callison19a,Hastings19a} can be made for why QW with short runtimes seems to perform well, a theoretical understanding of rapid quenches with time-dependent Hamiltonians, but far from the regime where the adiabatic theorem applies, is essentially missing. 

It has been recently shown numerically \citep{Brady2020a} that the optimal protocol for solving problems often involves an annealing step, as opposed to bang-bang controls where driver and problem Hamiltonians are not active simultaneously. Previous theoretical work based on the Pontryagin's minimum principle showed that optimal control patterns would always take the bang-bang form \citep{Yang17a}, but that these controls would sometimes require un-physically switching between the driver and problem Hamiltonians an infinite number of times in a finite time span. The work done by Brady et al \citep{Brady2020a} is restricted to cases with a finite number of ``bangs'' and found that in this more realistic setting protocols involving annealing may be superior.

We begin in section \ref{sec:mon_examples} with some numerical examples to illustrate the performance gains that can be obtained from well-chosen rapid quenches in quantum annealing.
This provides motivation to understand why rapid quenches work, and how to exploit the effects more systematically.
We then analyse the energy flow between different quantum states, altering
the expectation values of driver and problem terms in the Hamiltonian, as laid out in section \ref{sec:en_perspect}.
Next, we provide a general set of conditions (essentially requiring that quenches be monotonic) under which rapid quenches will preferentially seek out high quality solutions.
We augment this analysis by studying the transitions between different computational basis states, to deduce the level of dynamics which will occur, in section \ref{sec:ensure_dyn}, and, we apply our tools to different problem settings, including discussing the conditions for general optimisation problems to yield a significant level of dynamics.
Then, in section \ref{sec:s_heur} we show how the tools developed here can be used to construct heuristics for setting the parameters for continuous time quantum walks and rapid quenches.
Section \ref{sec:num_meth} provides details of our numerical methods, and we summarise and discuss our results in section \ref{sec:conc}.

\section{Rapid quench examples \label{sec:mon_examples}} %%%%%%%%%%%%%%%%%%%%%%%

To motivate our theoretical tools, we start with three illustrative examples showing the power of rapid quenches to solve problems.
For simplicity and concreteness, we focus on \textit{monotonic} quenches; that is, quenches for which the control parameter $\Gamma(t')\leq\Gamma(t)$ $\forall t'>t$.

\subsection{Two stage quantum walk}\label{ssec:two-stage-walk} %%%%%%%%%%%%%%%%

This is a minimal modification to the time-independent continuous time quantum walk.
It consists of two time-independent stages of evolution separated by an infinitely fast quench.
Because each stage is effectively a continuous time quantum walk, we refer to this as a \emph{two-stage} quantum walk.
\begin{figure}
\begin{centering}
\includegraphics[width=0.99\columnwidth]{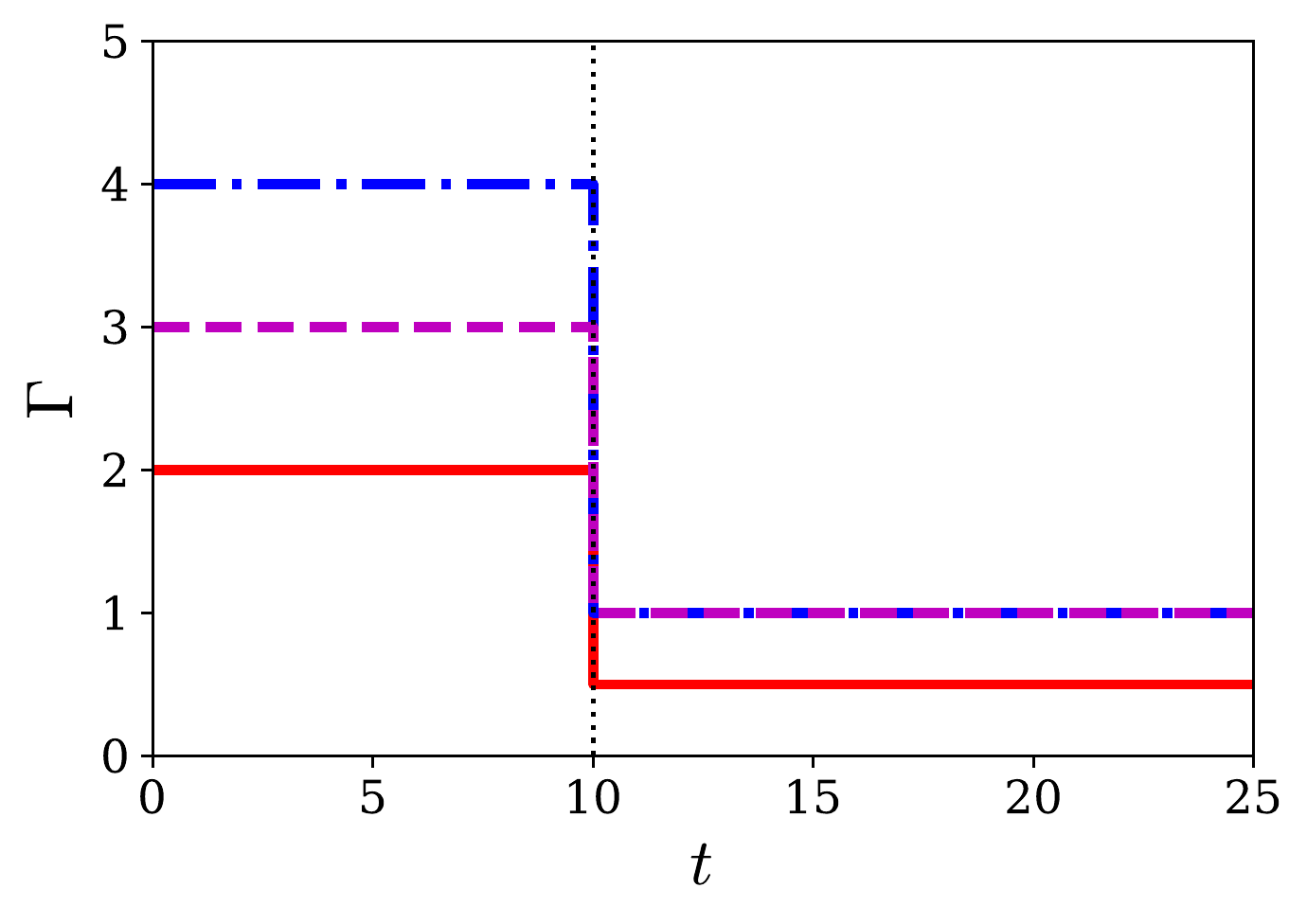}
\par
\caption{\label{fig:2_stage_schedule}The annealing schedule for the two stage quantum walks in Fig.~\ref{fig:2stage_simple} (red, solid lines), Fig.~\ref{fig:2stage_sk_qw} (blue, dot dashed lines), and Fig.~\ref{fig:2stage_bias_sk} (magenta, dashed lines). In all cases the step occurs at $t=10$ (dotted line).
}
\end{centering}
\end{figure}
We use a simple transverse field driver Hamiltonian
\begin{eqnarray}
H_\mathrm{drive} &=& n\openone - \sum_{j=1}^{n}\hat{X}_j, \label{eqn:transverse_field}
\end{eqnarray}
where $\openone$ is the identity operator and $\hat{X}_j$ is the Pauli $\hat{X}$ operator acting in qubit $j$.  Instead of using a constant control function ($\Gamma(t)=\gamma$), we use the time-dependent schedule
\begin{equation}
\Gamma(t) = \Bigg\{\begin{array}{ll}
\gamma_1 & 0<t<t_1 \\
\gamma_2 & t_1<t<(t_1+t_2)
\end{array},
\end{equation}
which consists of two consecutive evolution stages with two different time independent Hamiltonians.
Each of these stages is effectively a quantum walk, although the second stage uses non-standard starting conditions as its initial state is the final state of the first stage.
The standard initial state is the equal superposition of all basis states,
$\ket{\psi_0}=2^{-n/2}\sum_j\ket{j}$, chosen because it is the ground state of the driver Hamiltonian, and also represents our ignorance of which basis state is the solution to the problem.
The schedules we use for the two stage quantum walks are shown in Fig.~\ref{fig:2_stage_schedule} for each of our three examples.
\begin{figure}
\begin{centering}
\includegraphics[width=0.99\columnwidth]{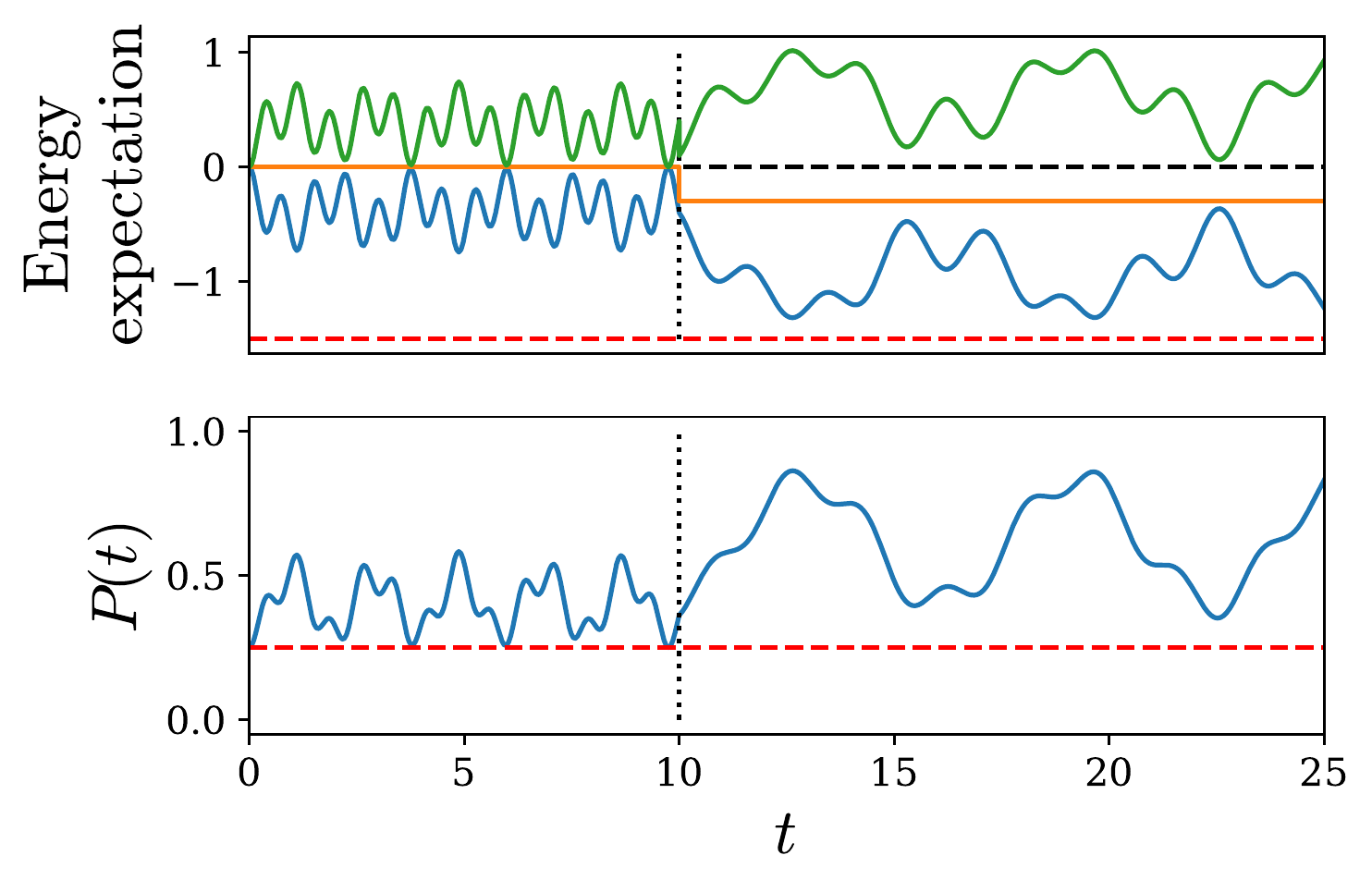}
\par
\caption{\label{fig:2stage_simple}
Two stage quantum walk using Hamiltonian in (\ref{eq:H_simple}) with $\gamma_1=2$ and $\gamma_2=\frac{1}{2}$.
The instantaneous quench occurs at time $t_1=10$ (vertical dotted line).
Top: energy expectation values $E_{\Gamma}=\Gamma\,\average{H_{\mathrm{drive}}}+\average{H_{\mathrm{prob}}}$ (gold), $\Gamma \,\average{H_{\mathrm{drive}}}$ (green), $\average{H_{\mathrm{prob}}}$ (blue).
Also shown (black, dashed) is a guide to the eye at $0$, and the minimum eigenvalue of $H_{\mathrm{prob}}$ (red, dashed).
Bottom: probability $P(t)$ of being in the ground state of $H_{\mathrm{prob}}$ at time $t$ (blue), probability of random guessing (red, dashed).
}
\end{centering}
\end{figure}

As discussed in \cite{callison19a}, a quantum walk can be understood from an energetic perspective according to a mechanism referred to there as the \textit{energy conservation mechanism}.  Being time-independent, quantum walks conserve the total energy of the system.
To show the effect of changing the hopping rate $\gamma$ part way through the walk, thus disrupting the energy conservation, our first example is a simple two qubit problem Hamiltonian
\begin{equation}
H_{\mathrm{prob}}^{(2Q)}=-\hat{Z}_1\hat{Z}_2-\frac{1}{2} \hat{Z}_1 \label{eq:H_simple},
\end{equation}
where $\hat{Z}_j$ is the Pauli $\hat{Z}$ operator acting on qubit $j$.
We start the system at $t=0$ in the state $\ket{\psi_0}=\frac{1}{2}(\ket{00}+\ket{01}+\ket{10}+\ket{11})$, the two qubit ground state of the driver Hamiltonian $H_{\mathrm{drive}}$ in (\ref{eqn:transverse_field}).
To simplify notation, we define $\average{H_{\mathrm{prob}}}_{\psi(t)}\equiv \sandwich{\psi(t)}{H_{\mathrm{prob}}}{\psi(t)}$, the instantaneous expectation value of the problem Hamiltonian with respect to the state $|\psi(t)\rangle $ at time $t$.  Likewise, $\average{H_{\mathrm{drive}}}_{\psi(t)}\equiv \sandwich{\psi(t)}{H_{\mathrm{drive}}}{\psi(t)}$ for the driver Hamiltonian.  We have the total energy $E_{\Gamma}(t)=\Gamma(t)\average{H_{\mathrm{drive}}}+\average{H_{\mathrm{prob}}}$.

Figure ~\ref{fig:2stage_simple} (top) shows that the expectation value $\langle H_\mathrm{drive} \rangle$ for the transverse field is zero initially ($t=0$).
As in \cite{callison19a}, the energy conservation mechanism then decreases the expectation value of the problem Hamiltonian at the expense of increasing the expectation value of the driver Hamiltonian.
When the instantaneous quench is performed, the problem Hamiltonian expectation value is unchanged, but the driver Hamiltonian expectation value (and therefore the total energy expectation value $E_\Gamma(t)$) is reduced.
As the minimum eigenvalue of $H_{\mathrm{drive}}$ is zero, the total energy expectation value $E_\Gamma(t)$ acts as an effective upper bound on $\average{H_{\mathrm{prob}}}_{\psi(t)}$.
The net effect is that, even if all of the energy stored in the transverse field were returned to the problem Hamiltonian, its expectation value would still be less than it was at the beginning of the algorithm.
What actually happens, however, is that the transverse field is able to capture even more of the energy, thereby reducing the problem Hamiltonian expectation value further, and increasing the average probability of finding the ground state, Fig.\ref{fig:2stage_simple} (bottom).

A more realistic problem is the Sherrington-Kirkpatrick spin-glass \citep{kirkpatrick1975solvable} ground-state problem investigated in \cite{callison19a}.
This has the problem Hamiltonian
\begin{eqnarray}\label{eqn:SKh}
    H_\mathrm{prob}^{(SK)} = -\frac{1}{2}\sum_{(a\neq b)=0}^{n-1}J_{ab}\hat{Z}_a\hat{Z}_b - \sum_{b=0}^{n-1}h_b\hat{Z}_b,
\end{eqnarray}
the couplings $J_{ab}$ and fields $h_b$ are drawn independently from the normal distribution $\mathcal{N}(0,\sigma^2)$ with mean $0$ and variance $\sigma^2$.

\begin{figure}
\begin{centering}
\includegraphics[width=0.99\columnwidth]{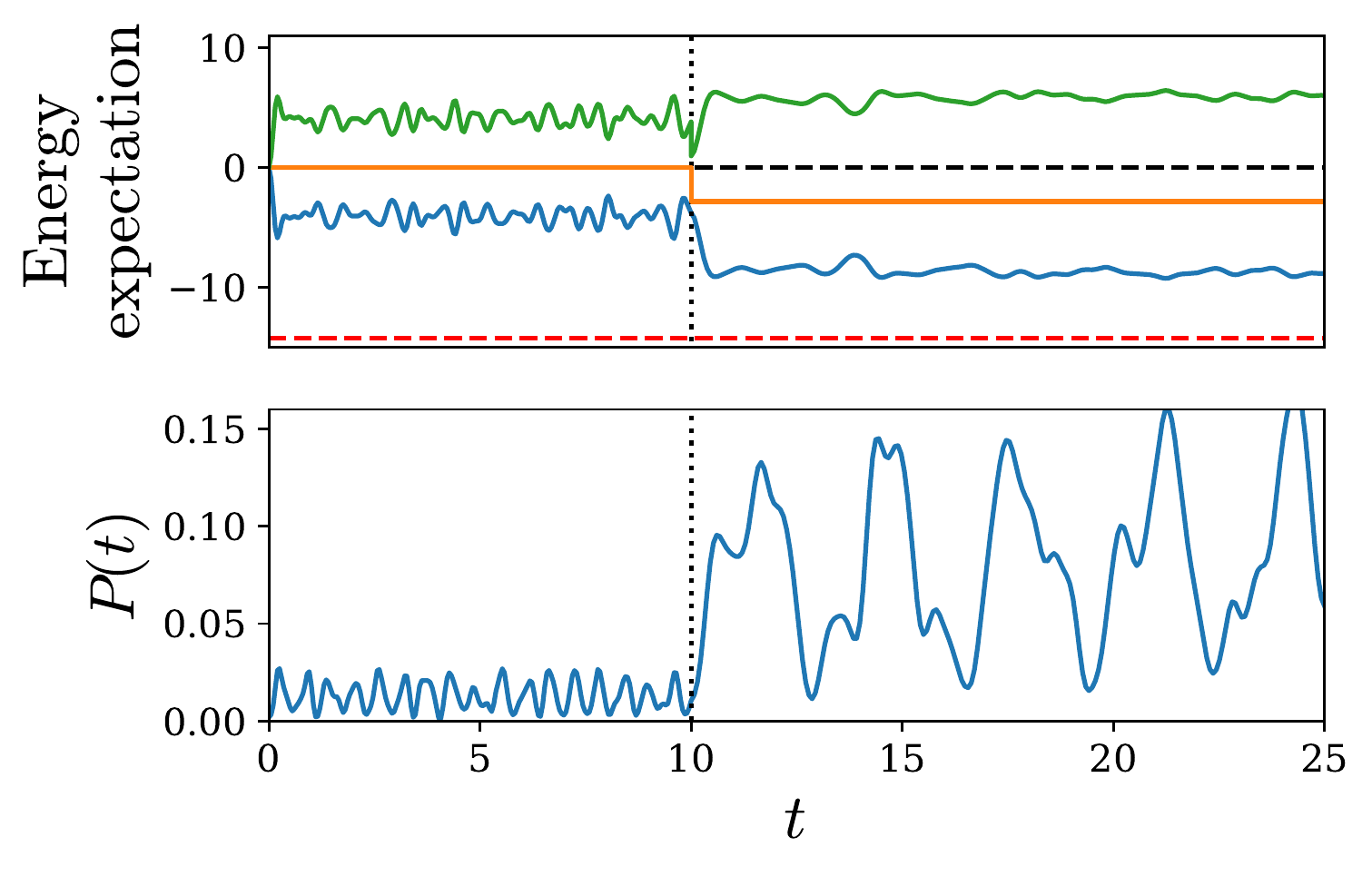}
\caption{\label{fig:2stage_sk_qw}
Two stage quantum walk on a $9$ qubit Sherrington-Kirkpatrick spin glass, ID code \textbf{ovcjhwbhtcpcvwicoxpdpvjzqojril} from the public repository in \cite{data_arch_SG} with $\gamma_1=4$ and $\gamma_2=1$.
The instantaneous quench occurs at time $t_1=10$, (vertical dotted line).
Top: energy expectations $E_{\Gamma}=\Gamma\,\average{H_{\mathrm{drive}}}_{\psi(t)}+\average{H_{\mathrm{prob}}}_{\psi(t)}$  (gold), $\Gamma \,\average{H_{\mathrm{drive}}}_{\psi(t)}$ (green),  $\average{H_{\mathrm{prob}}}_{\psi(t)}$ (blue). Also shown (black, dashed) is a guide to the eye at $0$, and the minimum eigenvalue of $H_{\mathrm{prob}}$ (red, dashed).
Bottom: probability $P(t)$ of being in the ground state of $H_{\mathrm{prob}}$ at time $t$ (blue).
}
\end{centering}
\end{figure}
Figure \ref{fig:2stage_sk_qw} shows a two stage quantum walk performed on a nine qubit Sherrington-Kirkpatrick Hamiltonian \footnote{%
We chose instance \textbf{ovcjhwbhtcpcvwicoxpdpvjzqojril} and used it throughout this paper for all single SK problem examples.}
from the public repository \cite{data_arch_SG} associated with \cite{callison19a}.
In the setting of this larger problem, the fluctuations after each stage of the quantum walk are smaller relative to the dynamical range than in the two qubit case, a very early sign of the approach to the thermodynamic limit.   Apart from this, the behaviour is qualitatively similar to the two qubit toy model $H_{\mathrm{prob}}^{(2Q)}$ of (\ref{eq:H_simple}) shown in Fig.~\ref{fig:2stage_simple}, and produces a significant increase in the probability of finding the ground state.

\subsection{Biased two stage quantum walk}\label{ssec:biased2QW} %%%%%%%%%%%%%%%%%%%%%

We introduce a biased driver Hamiltonian, similar to the one used in \cite{Duan2013a,Grass19a}.
We formulate our biased driver Hamiltonian slightly differently as
\begin{equation}
H_{\mathrm{bias}}(g,\theta)=n\,\openone-\sum_{i=1}^n\left(\cos(\theta)\hat{X}_i+g_i\,\sin(\theta)\hat{Z}_i\right), \label{eq:Hbias}
\end{equation}
where $g_i\in \{-1,1\}$ is a candidate (or guess) solution, and takes the value $1$ if the $i$th bit of the guess solution is $0$, and $-1$ if it is $1$.
The certainty of the guess is parametrized by $0\le \theta \le \frac{\pi}{2}$; if $\theta=0$ the guess goes unused and the driver reduces to a transverse field of (\ref{eqn:transverse_field}).
In the other extreme, if $\theta=\frac{\pi}{2}$, then the ground state of $H_{\mathrm{bias}}(g,\theta =\frac{\pi}{2})$ is the candidate solution and there are no dynamics.
The ground state of the biased driver Hamiltonian has zero energy for all allowed values of $\theta$ and $g$, and is a tensor product of spin states which are each anti-parallel to the fields in (\ref{eq:Hbias}); this state is used as the initial state.
For simplicity, in this example we only consider biasing toward the most optimal solution (i.e., correct guesses), and we use the same nine-qubit SK spin glass as in the previous subsection.

\begin{figure}
\begin{centering}
\includegraphics[width=0.99\columnwidth]{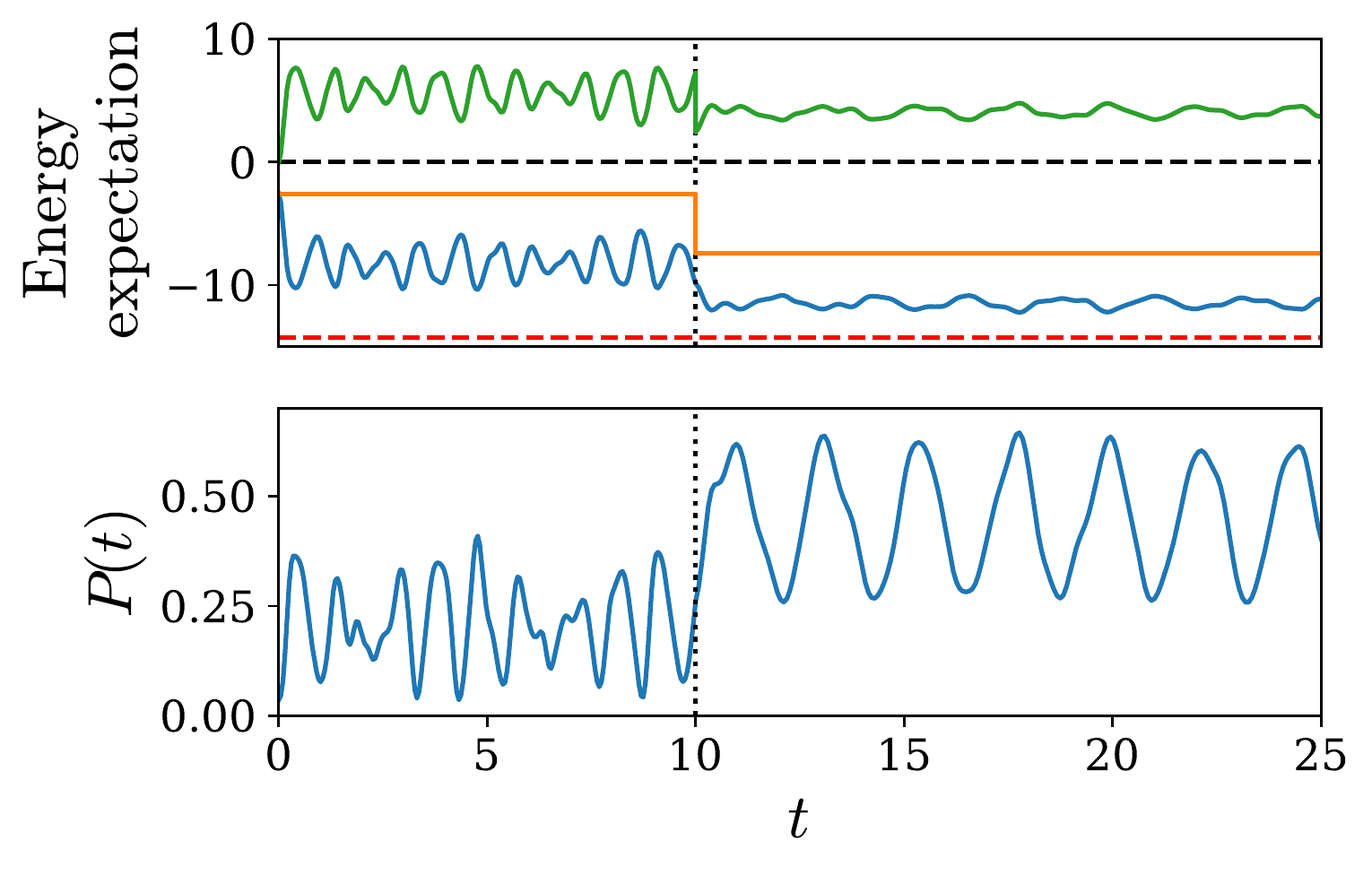}
\par
\caption{\label{fig:2stage_bias_sk}
Biased two stage quantum walk on a $9$ qubit Sherrington-Kirkpatrick spin glass, ID code \textbf{ovcjhwbhtcpcvwicoxpdpvjzqojril} from the public repository in \cite{data_arch_SG} with $\gamma_1=3$ and $\gamma_2=1$, using a biased driver (\ref{eq:Hbias}), biased towards the optimal solution of $H_{\mathrm{prob}}$ using $\theta=\frac{\pi}{8}$. The instantaneous quench occurs at time $t_1=10$ (vertical dotted line).
Top: energy expectations $E_\Gamma=\Gamma\,\average{H_{\mathrm{drive}}}_{\psi(t)}+\average{H_{\mathrm{prob}}}_{\psi(t)}$ (gold), $(\Gamma\,\average{H_{\mathrm{drive}}}_{\psi(t)}$ (green), $\average{H_{\mathrm{prob}}}_{\psi(t)}$ (blue). Also shown (black, dashed) is a guide to the eye at $0$, and the minimum eigenvalue of $H_{\mathrm{prob}}$ (red, dashed).
Bottom: probability $P(t)$ of being in the ground state of $H_{\mathrm{prob}}$ at time $t$. (blue).
}
\end{centering}
\end{figure}
As Fig.~\ref{fig:2stage_bias_sk} shows, the effect of biasing toward the optimal solution is to lower the initial values of $E_\Gamma$ and $\average{H_\mathrm{prob}}_{\psi(t)}$; biasing toward a well chosen guess effectively gives the algorithm a `head start' with respect to energy expectation values.
This is qualitatively similar to what happens at the beginning of the second stage of the two stage quantum walk, except that the driver energy $\average{H_{\mathrm{bias}}(g,\theta)}_{\psi(t)}$ starts at exactly zero, rather than having some initial energy left over from a previous stage.
The bias improves the initial stage success probability by a factor of ten compared with the unbiased walk in Fig.~\ref{fig:2stage_sk_qw}, while the second stage again provides a (further) factor of three improvement.
This biased two-stage quantum walk example provides proof-of-concept that the mechanism we describe can be leveraged on top of a biased search.
A thorough analysis of biased (single stage) quantum walks as a subroutine for hybrid quantum/classical computing is forthcoming \citep{Nita19a}.

\subsection{Pre-annealed quantum walk}\label{ssec:preanneal} %%%%%%%%%%%%%%%%%%%%%

Our final example is again in two stages, but this time the first stage is a quantum anneal, and the second stage is a quantum walk that starts from the point where the anneal stops.
The motivating intuition is that the initial time-dependent annealing stage will prepare an initial state for the quantum walk that has a lower average problem energy $\average{H_\mathrm{prob}}_{\psi(t)}$ than the usual uniform superposition state.
If performed too slowly, such a quench will put the system into its instantaneous ground state, by the adiabatic theorem of quantum mechanics, and there will be no quantum walk dynamics.
If performed too rapidly, the state will not evolve much during the anneal stage and the resulting quantum walk will be similar to one without a pre-annealing stage.
However, if the anneal is performed at an intermediate rate, it leads to significant quantum walk dynamics, starting from a lower problem Hamiltonian expectation value $\average{H_\mathrm{prob}}_{\psi(t)}$.

\begin{figure}
\begin{centering}
\includegraphics[width=0.99\columnwidth]{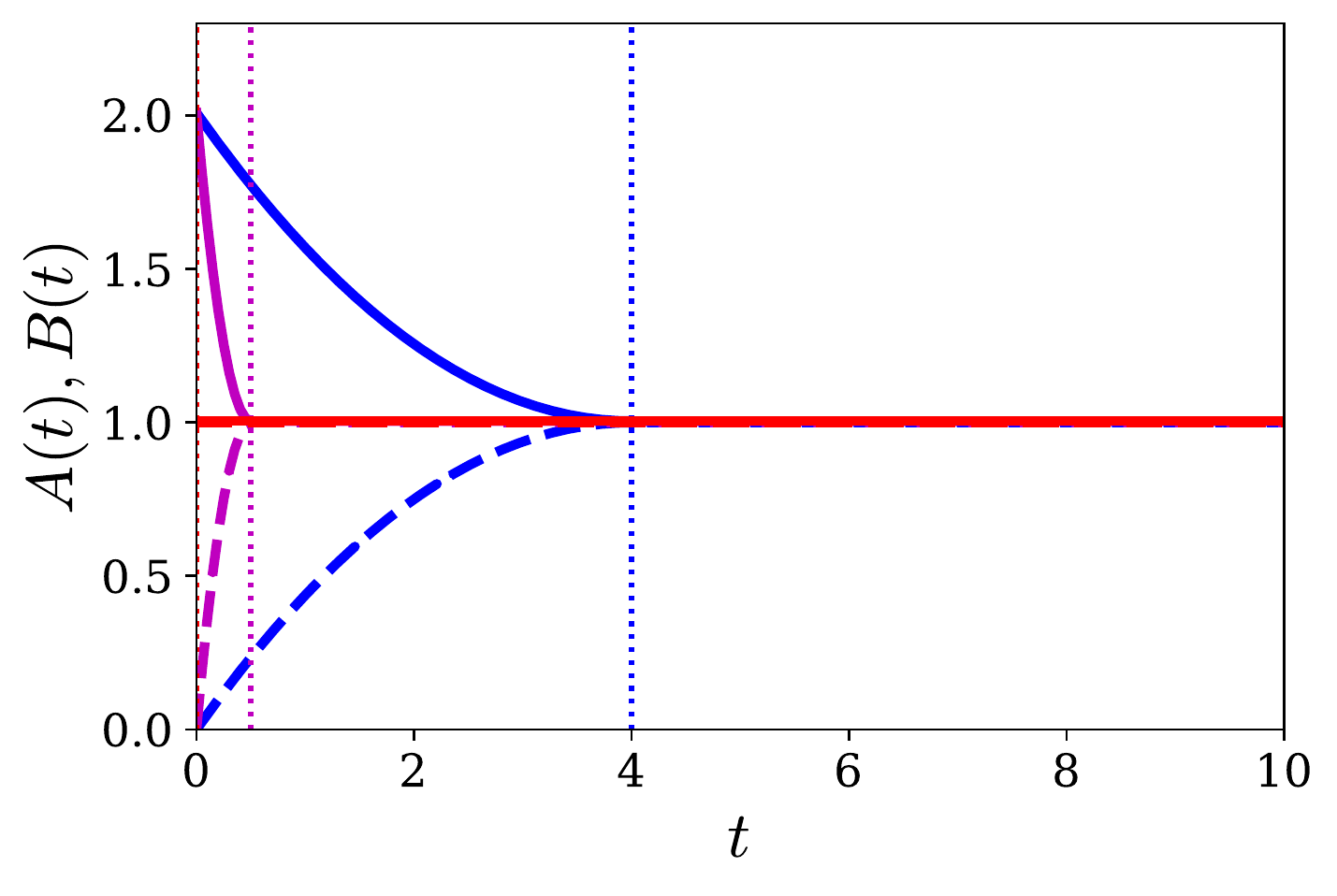}
\caption{\label{fig:preanneal_sched}
Schedule $A(t)$ (solid) and $B(t)$ (dashed) of a pre-annealed quantum walk using $\gamma\approx 1.004$ and $t_1=4$ (blue, vertical dotted line), $t_1=0.5$ (magenta, vertical dotted line), and $t_1=0$, (red, pure QW).}
\end{centering}
\end{figure}

Using the $H_{AB}$ parametrization defined in (\ref{eq:Hopt}), we consider pre-annealing with a quadratic schedule for a time $t_1$, and then a steady state quantum walk afterwards; specifically, we define the schedule
\begin{equation}
A(t) = \Bigg\{\begin{array}{ll}
\gamma[1+(\frac{t}{t_1}-1)^2] & 0\le t \le t_1 \\
\gamma & t_1<t\leq(t_1+t_2)
\end{array},
\end{equation}
\begin{equation}
B(t) = \Bigg\{\begin{array}{ll}
[1-(\frac{t}{t_1}-1)^2] & 0\le t \le t_1 \\
1 & t_1<t\leq(t_1+t_2)
\end{array}
\end{equation}
which is plotted in Fig.~\ref{fig:preanneal_sched} for the values of $t_1$ we use.

\begin{figure}
\begin{centering}
\includegraphics[width=0.99\columnwidth]{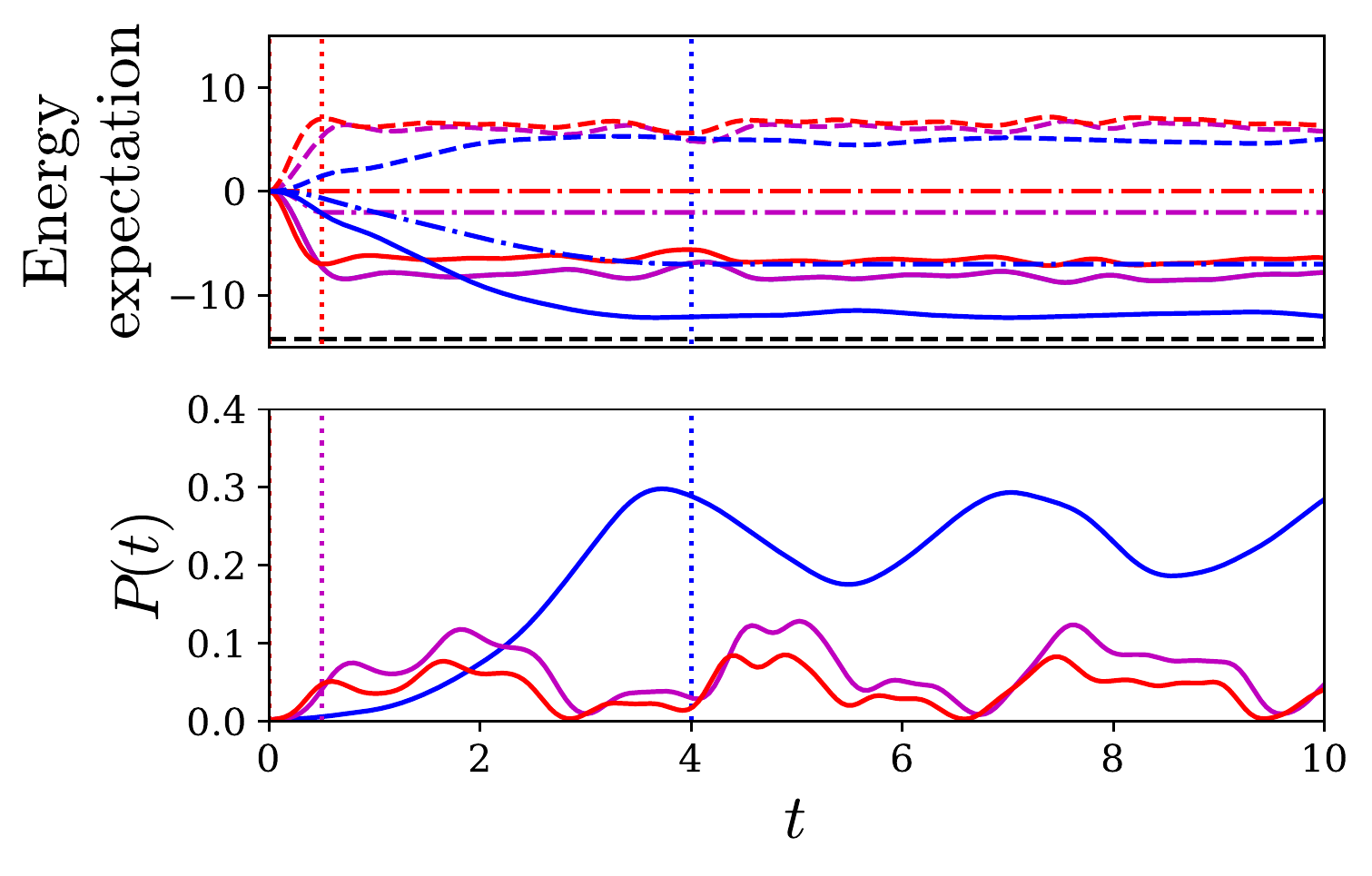}
\caption{\label{fig:preanneal_multi}Pre-anneal performed on a nine qubit Sherrington-Kirkpatrick spin glass, ID code \textbf{ovcjhwbhtcpcvwicoxpdpvjzqojril} from \cite{data_arch_SG}, for pre-anneal times $t_1=4$ (blue), $t_1=0.5$ (magenta), $t_1=0$ (red), i.e., pure quantum walk. Dotted lines show when the pre-anneal ends. Top: Expectation values $E_\Gamma(t)=\frac{A}{B}\average{H_{\mathrm{drive}}}_{\psi(t)}+\average{H_{\mathrm{prob}}}_{\psi(t)}$ (dot-dashed), $\average{H_{\mathrm{prob}}}_{\psi(t)}$ (solid), $\frac{A}{B}\average{H_{\mathrm{drive}}}_{\psi(t)}$ (dashed, colour). The black dashed line indicates the minimum eigenvalue of  $H_{\mathrm{prob}}$. Bottom: success probability $P(t)$ to be in the lowest eigenstate of $H_{\mathrm{prob}}$ at time $t$.}
\end{centering}
\end{figure}

Using the same nine-qubit SK problem as before, with its optimal $\gamma$ value of approximately $1.004$, the results for three different values of $t_1$ are shown in Fig.~\ref{fig:preanneal_multi}.
Pre-annealing both decreases the average problem expectation value $\langle H_\mathrm{prob}\rangle_{\psi(t)}$ and increases the success probability, but causes the peak values to be reached more slowly.
In the longest pre-anneal with $t_1=4$, the success probability undergoes small amplitude, approximately sinusoidal, oscillations suggesting that the dynamics are dominated by a two level subspace. 
For $t_1=0.5$ and $t_1=0$, the oscillations are less structured, indicating that more than two energy levels are playing a non-trivial role in the dynamics.

The increases in the success probability seen in Fig.~\ref{fig:preanneal_multi} are relatively modest for this example. 
To determine the typical improvement in success probability due to pre-annealing, we use all $10,000$ Sherrington-Kirkpatrick instances from \cite{data_arch_SG} at each size from $n=5$ to $n=11$ and compare the quantum walk success probability averaged over the quantum walk stage using $20$ different linearly spaced pre-annealing times up to $t_1=4$.
\begin{figure}
\begin{centering}
\includegraphics[width=0.99\columnwidth]{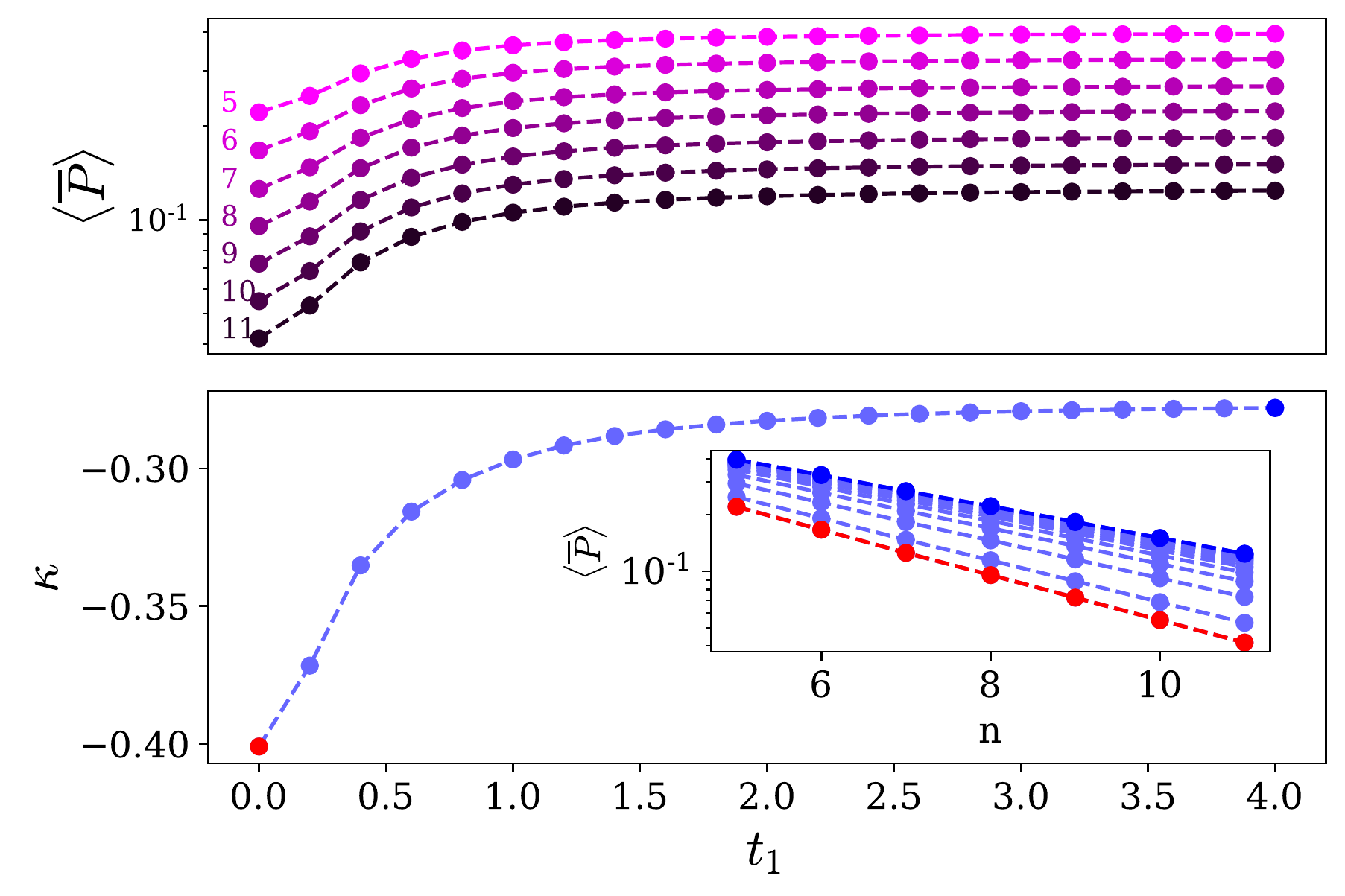}
\caption{\label{fig:preanneal_scaling}
Top: success probability $\langle \overline{P} \rangle$ for $n=5$ to $n=11$ for $21$ different linearly spaced pre-anneal times from $t_1=0$ to $t_1=4$,  darker magenta colour indicates higher $n$. All data are averaged over all $10,000$ Sherrington-Kirkpatrick instances from \cite{data_arch_SG} at each size. 
Bottom: Scaling exponent $\kappa$ for a model where $p_{\mathrm{success}}\propto 2^{\kappa n}$ extracted from the linear fit on log-linear axes for different pre-annealing times in the inset. 
Inset: Scaling of success probability versus $n$, for the same $t_1$ values, with $t_1=0$ in red, and $t_1=4$ in dark blue (same colour coding as the bottom main figure). 
}
\end{centering}
\end{figure}
In Fig.~\ref{fig:preanneal_scaling} (top), we see that the success probability increases with pre-anneal time, up to a plateau, and the relative effect of pre-annealing becomes larger as $n$ increases.
To quantify this effect, we calculate the scaling exponent at each pre-annealing time by fitting a linear model on log-linear axes.
We find a scaling exponent $\kappa$ such that the success probability $p_{\mathrm{success}}\propto 2^{\kappa n}$.
The fitted values of $\kappa$ are plotted in  Fig.~\ref{fig:preanneal_scaling} (bottom).
As the inset of Fig.~\ref{fig:preanneal_scaling} (bottom) shows, the success probability is modelled well by a simple exponential function, as in \cite{callison19a}.
We find that pre-annealing significantly improves the scaling from $\kappa = -0.418$ for a pure quantum walk, in agreement with \cite{callison19a}, to a maximum of $\kappa \approx -0.278$.
It is, of course, an open question whether or not this scaling will continue to problem sizes which are of practical interest, but the lack of visible finite size effects in  Fig.~\ref{fig:preanneal_scaling} suggests that it might.
Since very fast quenches can be experimentally challenging to implement, although methods are being explored \citep{lanting_AQC2018}, determining the effects of quenching at a finite rate is of practical importance.  Our results show that such quenches are potentially a better strategy than trying to speed up or slow down to approach QW or adiabatic extremes.

\section{Energy redistribution mechanism \label{sec:en_perspect}} %%%%%%%%%%%%%%%%%%%%%%%%

In all the examples in section \ref{sec:mon_examples}, we observe that
the total energy expectation value $E_\Gamma(t)$ never increases during a rapid quench, and that $E_\Gamma(t)$ serves as an upper-bound to the problem expectation value $\langle H_\mathrm{prob}\rangle_{\psi(t)}$,
assuming that the groundstate of $H_{\mathrm{drive}}$ is arranged to be at zero energy (the identity term in (\ref{eqn:transverse_field}) ensures this).
In this section, we formalise these observations into a mechanism that we refer to as the \emph{energy redistribution mechanism}.
Our analysis extends the energy conservation mechanism described in \cite{callison19a} and recapped in Appendix \ref{app:enconsmech} (similar arguments are also made by Hastings in \cite{Hastings19a}) to quenches where the Hamiltonian is not time invariant, and therefore total energy is not conserved.

Consider a closed system quantum annealing schedule on a system with a Hamiltonian $H(t)$ defined by (\ref{eq:Hopt_Gamma}):
\begin{equation}
H(t)=\Gamma(t)\,H_{\mathrm{drive}}+H_{\mathrm{prob}}.
\end{equation}
We show that (for duration $t_f \ge 0$) the energy expectation value with respect to the problem Hamiltonian at the end is never higher than at the initial time $t=0$,
\begin{eqnarray}
\sandwich{\psi(t_f)}{H_{\mathrm{prob}}}{\psi(t_f)} &\le& \sandwich{\psi(0)}{H_{\mathrm{prob}}}{\psi(0)}, \label{eqn:Hp_decreases}
\end{eqnarray}
provided the following conditions are satisfied:
\begin{enumerate}
\item (\emph{initial ground state}) the initial state $\ket{\psi(t=0)}$ is a ground state of the driver Hamiltonian $H_{\mathrm{drive}}$ \label{condition_initial_ground_state}
\item (\emph{positivity}) the control function is non-negative: $\Gamma(t)\ge 0$ $\forall t$ \label{condition_nonnegative_Gamma}
\item  (\emph{monotonicity}) the control function is monotonically decreasing: $\Gamma(t) \ge \Gamma(t')$ $\forall t'>t$ \label{condition_monotonic_Gamma}
\end{enumerate}
Condition \ref{condition_initial_ground_state}.~is simply that the system is initially prepared in the ground state of the driver Hamiltonian.
This condition is necessary for AQC, and is also standard for QW.
Condition \ref{condition_nonnegative_Gamma}.~prevents pathological behaviour where the driver spectrum is effectively inverted by taking negative values of the control function $\Gamma(t)$.
This condition is satisfied in all traditional AQC and QW settings.
Condition \ref{condition_monotonic_Gamma}.~is that the quench is \emph{monotonic}; this condition excludes methods such as reverse annealing, both the dissipatively driven form proposed in \cite{chancellor17b} and implemented on D-Wave devices \citep{reverse_anneling_whitepaper}, and the similar coherent method proposed in \cite{Perdomo-Ortiz11} which is sometimes also referred to as reverse annealing.
The biased driver Hamiltonian proposed in \cite{Duan2013a,Grass19a} is compatible with  condition \ref{condition_monotonic_Gamma}.
Our results do not rely on the adiabatic theorem and the control function $\Gamma(t)$ does \emph{not} need to be a continuous function.

Without loss of generality, the driver Hamiltonian $H_\mathrm{drive}$ can be chosen such that its ground-state eigenvalue (and hence its expectation value with the initial state) is zero $\sandwich{\psi(0)}{H_{\mathrm{drive}}}{\psi(0)}=0$. 
In other words, we impose \textit{semidefiniteness} on $H_{\mathrm{drive}}$ by defining its ground state $\ket{\psi(0)}$ to have eigenvalue 0.
Let 
\begin{eqnarray}
E_\Gamma(t) &=&\sandwich{\psi(t)}{H_\Gamma(t)}{\psi(t)}\label{eq:totalenergyexpect}
\end{eqnarray}
be the expectation value of the energy at time $t$.
Then, it follows immediately from condition \ref{condition_initial_ground_state}.~that, at time $t=0$,
\begin{eqnarray}
E_\Gamma(0) &=& \sandwich{\psi(0)}{H_\mathrm{prob}}{\psi(0)}.\label{eqn:energy_redist_part1}
\end{eqnarray}
Furthermore, it follows from conditions \ref{condition_initial_ground_state}.~and \ref{condition_nonnegative_Gamma}.~that, at any later time $t>0$,
\begin{eqnarray}
E_\Gamma(t) &\geq& \sandwich{\psi(t)}{H_\mathrm{prob}}{\psi(t)},\label{eqn:energy_redist_part2}
\end{eqnarray}
since $\sandwich{\psi(t)}{H_\mathrm{drive}}{\psi(t)}\ge\sandwich{\psi(0)}{H_\mathrm{drive}}{\psi(0)}=0$ can only increase from the ground state initial energy.

It can be shown that the energy expectation value $E_\Gamma (t)$ defined in (\ref{eq:totalenergyexpect}) decreases monotonically in time; that is 
\begin{eqnarray}
E_\Gamma(t') \leq E_\Gamma(t)\,\forall t,t'\,: t'>t.  \label{eqn:energy_redist_part3}
\end{eqnarray}
To see this, consider the discretized approximation to the evolution
\begin{equation}
\ket{\psi_k^{(q)}}=\mathcal{T}\prod_{k'=k}^1 \exp(-iH_\Gamma(\frac{k't_f}{q})\frac{t_f}{q})\ket{\psi(0)},
\end{equation}
for $1\leq k \leq q$ and where the symbol $\mathcal{T}$ is added to emphasise that the time order of the product must be preserved, since the Hamiltonians at different times are non-commuting.
This discretized approximation becomes exact in the limit $q\rightarrow\infty$.
The evolution of a quantum system under the time dependent Hamiltonian given in (\ref{eq:Hopt}) from time $t=0$ to time $t=t_f$ from the initial state $\ket{\psi(0)}$ is broken down as follows: The initial state is evolved under the constant Hamiltonian $H(\frac{t_f}{q})$ for time $\frac{t_f}{q}$ to produce a state $\ket{\psi_1^{(q)}}$ which then evolves under the constant Hamiltonian $H(2\frac{t_f}{q})$ for time $\frac{t_f}{q}$ and so on, until a final state $\ket{\psi_q^{(q)}}$ is reached. Then, in the limit, $\ket{\psi(t_f)}=\lim_{q\rightarrow \infty}\ket{\psi_q^{(q)}}$. This kind of discretization can be thought of as an extension of the Suzuki-Trotter decomposition \citep{trotter59a,suzuki93a} and is therefore sometimes informally referred to as Trotterization.
In the same manner, we can define a discretized version of the energy expectation value as
\begin{eqnarray}
E_{\Gamma,k}^{(q)} &=& \Gamma\left(\frac{k't_f}{q}\right)\sandwich{\psi_k^{(q)}}{H_{\mathrm{drive}}}{\psi_k^{(q)}} + \nonumber\\
& &\sandwich{\psi_k^{(q)}}{H_{\mathrm{prob}}}{\psi_k^{(q)}},
\end{eqnarray}

During each time-independent evolution step, the energy expectation value $E_{\Gamma,k}^{(q)}$ is conserved.
Furthermore, since by definition $H_{\mathrm{drive}}$ is positive semidefinite and $\Gamma(\frac{(k+1)t_f}{q}) \leq \Gamma(\frac{kt_f}{q})$ (by conditions \ref{condition_nonnegative_Gamma}.~and \ref{condition_monotonic_Gamma}.), it follows that
\begin{equation}
E_{\Gamma,k+1}^{(q)} \leq E_{\Gamma,k}^{(q)}.
\end{equation}
Repeated application of this inequality results in the more useful inequality
\begin{equation}
E_{\Gamma,q}^{(q)} \leq E_{\Gamma,1}^{(q)}.
\end{equation}
Since $E_{\Gamma,1}^{(q)}$ is the energy during the whole of the first evolution step, it follows that
\begin{equation}
E_{\Gamma,1}^{(q)} = E_\Gamma(t=0).
\end{equation}
Furthermore, we have that
\begin{equation}
\lim_{q\rightarrow\infty} E_{\Gamma,q}^{(q)} = E_\Gamma(t_f).
\end{equation}
which means
\begin{equation}
E_\Gamma(t_f) \leq E_\Gamma(0) .
\end{equation}
Since this equation holds for all $t_f>0$, we have shown that $E_\Gamma(t)$ monotonically decreases with $t$, and (\ref{eqn:energy_redist_part3}) is proven.

Taken together, the statements in (\ref{eqn:energy_redist_part1}), (\ref{eqn:energy_redist_part2}) and (\ref{eqn:energy_redist_part3}) imply
\begin{eqnarray}
\sandwich{\psi(t=0)}{H_{\mathrm{prob}}}{\psi(t=0)} \ge \nonumber\\  \sandwich{\psi(t=t_f)}{H_{\mathrm{prob}}}{\psi(t=t_f)}, \label{eqn:quench_beats_guessing}
\end{eqnarray}
for final time $t_f$.
In other words, the energy expectation with respect to the problem Hamiltonian can only decrease compared with the initial state.
If the energy expectation of the problem Hamiltonian decreases, then the probability of measuring low energy states increases.

The result in (\ref{eqn:quench_beats_guessing}) holds for quenches, parameterized with the single control function $\Gamma(t)$, in the form of (\ref{eq:Hopt_Gamma}).
However, since the control function $\Gamma(t)$ is identified with the ratio $\nicefrac{A(t)}{B(t)}$ of control functions for quenches in the form of (\ref{eq:Hopt}), the result in (\ref{eqn:quench_beats_guessing}) follows automatically for quenches in $A(t)$, $B(t)$ form, except for when $B(0)=0$, when $\Gamma(0)$ is not well-defined.
In Appendix \ref{ssec:divergence_of_gamma}, we extend to the case where $B(0)=0$, with the additional condition that the driver Hamiltonian $H_\mathrm{drive}$ has a finite gap between its ground and first-excited manifolds (which is automatically true for all Hamiltonians on Hilbert spaces of finite dimension).

The key result is that, for quenches where the control function $\Gamma(t)$ decreases monotonically, the energy expectation value of the problem Hamiltonian $H_{\mathrm{prob}}$ cannot be higher than its initial value. Put another way, on average, a monotonic quench can never perform worse than random guessing.  This result is important for two reasons. Firstly, although not being harmful to average solution quality is a rather weak statement, it applies very generally to a broad class of algorithms. Secondly, and more importantly, this result can be built upon to determine control functions that can provide a significant improvement, which is important for algorithm design. To do this, we need to combine the result in this section with criteria for when the transfer of amplitude between computational basis states will be significant, which we obtain in the next section.

\section{Ensuring significant dynamics\label{sec:ensure_dyn}} %%%%%%%%%%%%%%%%%

In section \ref{sec:mon_examples}, we showed examples of a quantum quench giving significantly better performance than pure quantum walks.
In this section, we consider theoretically how a significant improvement can occur.
We know from section \ref{sec:en_perspect} that dynamics will never be detrimental; this means that, if dynamics occur, in general it will be beneficial. 
What remains is to determine the circumstances in which significant dynamics will occur.

\subsection{Quantifying the strength of short-time dynamics}\label{ssec:quantify_dyn} %%%%%%%%%%%%%%

In the analytical solutions for unstructured search in a continuous-time setting \citep{roland02a,childs03a}, the method involves analysing the dynamics in a two dimensional subspace.  
To obtain significant dynamics in this setting, the hopping rate $\gamma$ or schedule functions $A(t)$, $B(t)$ must carefully balance the relative strengths of the driver and problem Hamiltonians, such that the off-diagonal terms in the two dimensional subspace are maximized.  
Motivated by this, but being interested in shorter timescales, we instead investigate local subspaces spanned by a pair of basis states.
To analyse whether significant dynamics will occur, we perform a similar analysis to characterise how strong the transitions are to locally redistribute amplitude. If
these are large, for most of the transitions mediated by the driver, then the
system will generate a high level of dynamics on a short timescale; otherwise, it will
not, although dynamics may still occur on longer timescales.

As we want a measure of dynamics that can be efficiently estimated at all sizes, we analyse individual pairs of computational basis states connected by the driver, to determine whether significant transfer occurs between them, assuming the rest of the system remains in its initial state.
Note that, for classical problems in the setting we are considering, the problem Hamiltonian is diagonal in the computational basis, hence all of its subspaces are, too.
Consider two basis states $\ket{j}$ and $\ket{k}$ connected by the driver, i.e., $\sandwich{j}{H_{\mathrm{drive}}}{k}\ne 0$, and define an effective two-level system Hamiltonian
\begin{equation}
H_\Gamma^{(jk)}(t) = \Gamma(t) H_\mathrm{drive}^{(jk)} + H_\mathrm{prob}^{(jk)}\label{eq:H_sub_maintext}
\end{equation}
with the \emph{local problem Hamiltonian} $H_\mathrm{prob}^{(jk)}$ defined as 
\begin{equation}
H_\mathrm{prob}^{(jk)}=\left( \begin{array}{cc} E^{(j)} & 0 \\  0 & E^{(k)} \end{array} \right), \label{eq:Hp_sub_maintext}
\end{equation}
where $E^{(j)}=\sandwich{j}{H_{\mathrm{prob}}}{j}$ is the energy of computational basis state $\ket{j}$ with respect to the problem Hamiltonian (similarly for $k$), and with the \emph{local driver Hamiltonian} $H_\mathrm{driver}^{(jk)}$ defined as 
\begin{equation}
H_\mathrm{drive}^{(jk)}=\left( \begin{array}{cc} \sandwich{j}{H_\mathrm{drive}}{j} & \sandwich{j}{H_\mathrm{drive}}{k} \\  \sandwich{k}{H_\mathrm{drive}}{j} & \sandwich{k}{H_\mathrm{drive}}{k} \end{array} \right). \label{eq:Hd_sub_maintext}
\end{equation}
The extent to which the local subspace Hamiltonian $H_\Gamma^{(jk)}(t)$ can transfer amplitude between the basis states $\ket{j}$ and $\ket{k}$ can be characterised by comparing the off-diagonal energy scale to the diagonal one.
Define a local \textit{transfer coefficient}, which takes values $0 \le T^{(jk)}\le 1$, as
\begin{eqnarray}
T^{(jk)} & = & \mathcal{R}\Big[\Gamma(t)H_\mathrm{drive}^{(jk)}, H_\mathrm{prob}^{(jk)}\Big]\\
&\equiv& \frac{2\Gamma(t)|\sandwich{k}{H_\mathrm{drive}}{j}|}{2\Gamma(t)|\sandwich{k}{H_\mathrm{drive}}{j}| + |\Delta_{jk}|}. \label{eq:T_def}
\end{eqnarray}
where 
\begin{eqnarray}
\Delta_{jk} = \left\{\Gamma(t)\sandwich{j}{H_\mathrm{drive}}{j} + E^{(j)}\right\} - \nonumber\\
\left\{\Gamma(t)\sandwich{k}{H_\mathrm{drive}}{k} + E^{(k)}\right\}
\end{eqnarray}
is the difference between the diagonal elements in the diagonal basis of the problem Hamiltonian.

Similarly, as implied by the energy redistribution mechanism described in section \ref{sec:en_perspect}, transfer between driver eigenstates is also important. 
To capture this, we define a local \textit{driver coefficient}
$D^{(jk)}$ by transforming the local subspace Hamiltonian $H_\Gamma^{(jk)}(t)$ into the diagonal basis of the local driver Hamiltonian $H_\mathrm{drive}^{(jk)}$ and writing a similar expression to (\ref{eq:T_def}).
That is, 
\begin{eqnarray}
 D^{(jk)} & = & \mathcal{R}\/\Big[U^{(jk)\dagger}H_\mathrm{prob}^{(jk)}U^{(jk)}, \nonumber \\ 
& & \mbox{\ \ \ \ }\Gamma(t) U^{(jk)\dagger}H_\mathrm{drive}^{(jk)}U^{(jk)}\Big],
\end{eqnarray}
where $U^{(jk)}$ is a unitary such that $U^{(jk)\dagger}H^{(jk)}_\mathrm{drive}U^{(jk)}$ is diagonal.

It is easily shown that, for unbiased drivers such as (\ref{eqn:transverse_field}),
the local driver coefficient $D^{(jk)}$ and local transfer coefficient $T^{(jk)}$ are related by $D^{(jk)}=1-T^{(jk)}$.
This makes it clear there is a trade off between the two quantities to obtain significant dynamics under the combined Hamiltonian.
We quantify the overall level of amplitude transfer we expect by the product of the transfer and driver coefficients $T^{(jk)}$ and $D^{(jk)}$, which we call the dynamic coefficient,
\begin{equation}
\mathrm{Dyn}^{(jk)}=T^{(jk)}D^{(jk)}.\label{eq:Dyn_def}
\end{equation}
For unbiased drivers, since $D^{(jk)}=1-T^{(jk)}$, and $0\le D^{(jk)},T^{(jk)}\le 1$, it follows that $\mathrm{Dyn}^{(jk)}$ satisfies $0\le \mathrm{Dyn}^{(jk)}\le 0.25$.

The dynamic coefficient $\mathrm{Dyn}^{(jk)}$ captures the level of algorithmically useful local dynamics experienced by the system. 
In particular, if $\Gamma \gg 1$, then the driver Hamiltonian dominates and the problem Hamiltonian $H_\mathrm{prob}$ will have little effect on the dynamics of the system.
Since the initial state is the ground state of the driver Hamiltonian, the dynamics are driven by the much smaller problem Hamiltonian on short timescales.
This limit is captured by the dynamical coefficient, as $D^{(jk)}\approx 0$, and hence $\mathrm{Dyn}^{(jk)}\approx 0$.
In the opposite extreme, if $\Gamma\ll 1$, then the problem Hamiltonian dominates, but since it is diagonal, the dynamics will consist almost entirely of phase rotations in the computational basis, and the amplitudes will change very little.
This limit is captured by the transfer coefficient, as $T^{(jk)}\approx 0$, and hence $\mathrm{Dyn}^{(jk)}\approx 0$.

To characterise the level of dynamics in the entire system, we can simply take a mean value of $\mathrm{Dyn}^{(jk)}$ over the values of $j$ and $k$ which correspond to a non-zero off diagonal element in $H_{\mathrm{drive}}$. 
That is, we define the average dynamic coefficient
\begin{eqnarray}
\overline{\mathrm{Dyn}} &=& \langle \mathrm{Dyn}^{(jk)} \rangle_{jk}\label{eq:dynbar_def}
\end{eqnarray}
where $\langle \, \cdot \, \rangle_{jk}$ represents the mean over all pairs of computational basis states $j,k$ connected by the driver Hamilton $H_\mathrm{drive}$.
Although (\ref{eq:dynbar_def}) cannot be exactly calculated efficiently, it should in general be possible to approximate it efficiently (up to additive error) by sampling.
This follows from the fact that the values of $\mathrm{Dyn}^{(jk)}$ are bounded $0\le \mathrm{Dyn}^{(jk)}\le 0.25$, and therefore the error $\delta\overline{\mathrm{Dyn}}$ can be reduced to the range this value can take, multiplied by the statistical noise in the sample, which scales as the square root of the number of samples, i.e.,
\begin{equation}\label{eq:deltaDynbar}
    \delta\overline{\mathrm{Dyn}} \sim \frac{0.25}{N^{\nicefrac{1}{2}}_{\mathrm{samples}}}.
\end{equation}

As the average dynamic coefficient $\overline{\mathrm{Dyn}}$ is calculated by considering only those states that are directly connected by the driver Hamiltonian $H_\mathrm{drive}$, it naturally captures only the the fastest quantum dynamics that are present in the system. 
For example, in the case of the transverse field driver from (\ref{eqn:transverse_field}), $\overline{\mathrm{Dyn}}$ only depends on transitions between states that differ by a single bit-flip, which will typically be happening much faster than those that involve two or more bit flips.

\begin{figure}
  \subfigure[]{\includegraphics[width=0.99\columnwidth]{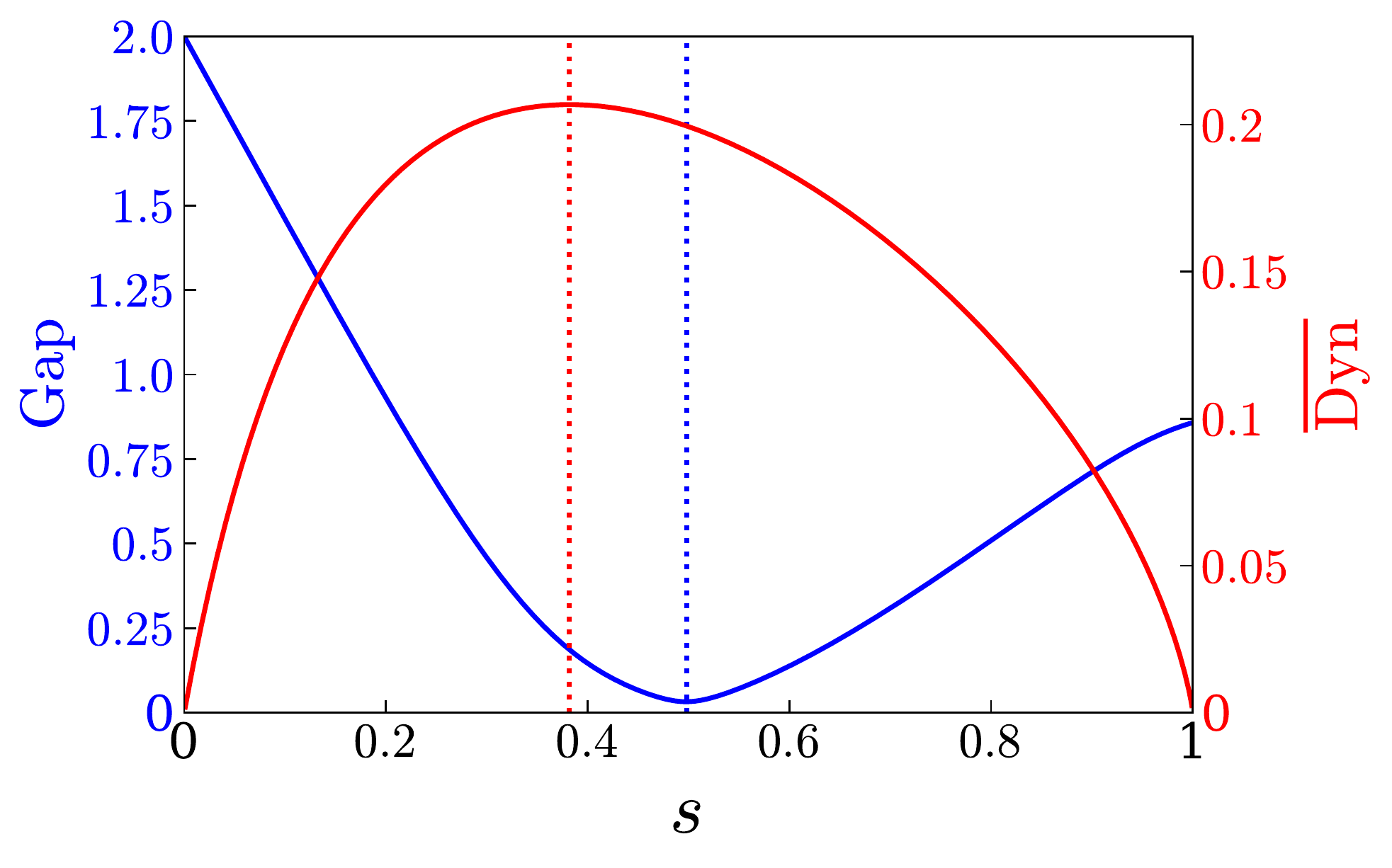}\label{fig:SKexample_gapdyn_sparam}}
  \subfigure[]{\includegraphics[width=0.99\columnwidth]{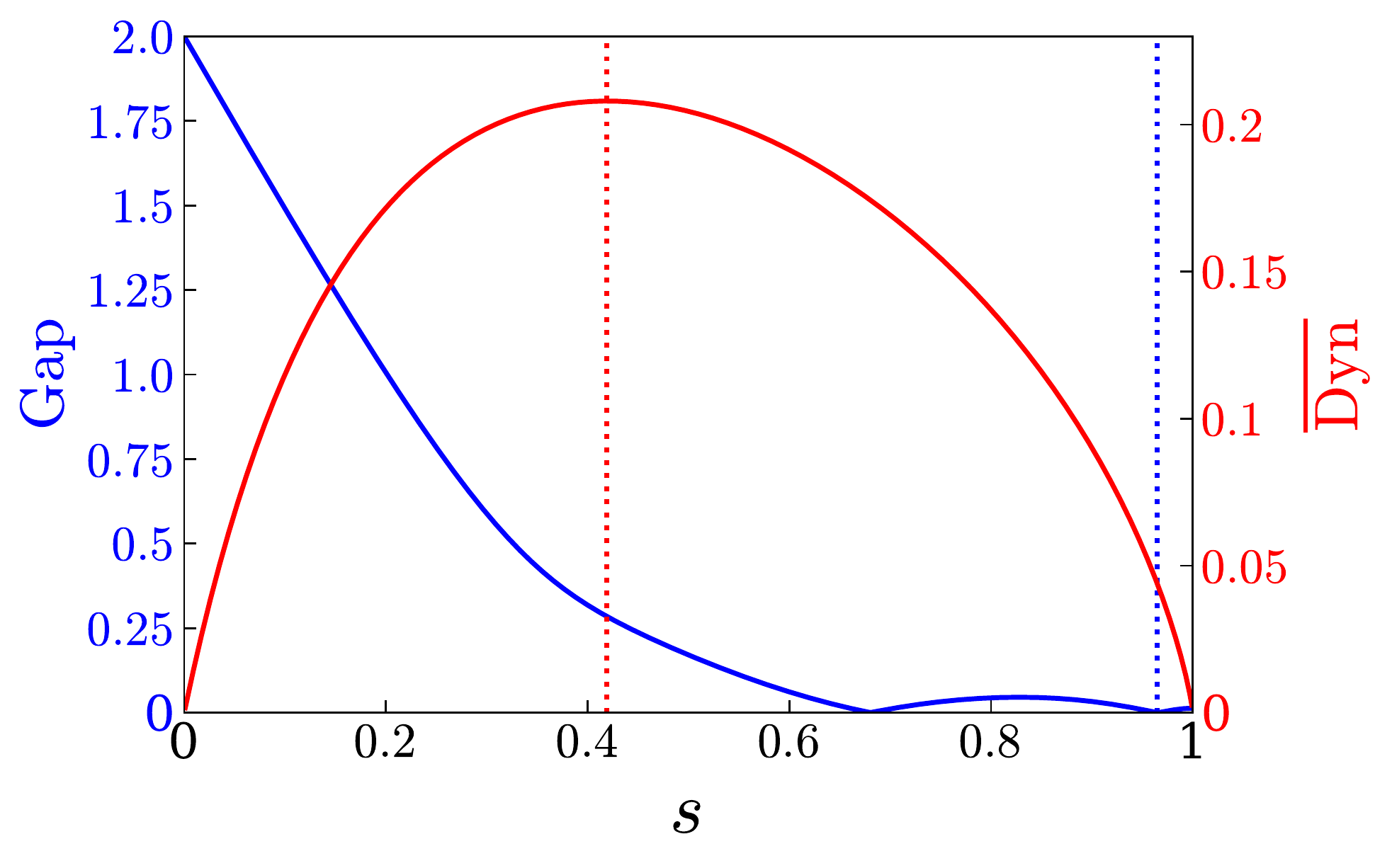}\label{fig:SKexample_gapdyn_sparam_smallest}}  
  \caption{Comparison, for two different $n=9$ SK instances, between the average dynamic coefficient $\overline{Dyn}$ (red solid line, right axis) with the gap between the ground- and first excited-state (blue solid line, left axis). 
  The quantities are plotted against the schedule parameter $s$ in the AQC-like paramaterization $A(t)=1-s(t), B(t)=s(t)$.
  The maximum of $\overline{\mathrm{Dyn}}$ and the minimum gap are indicated by the red and blue dotted lines respectively.
  Top: the instance with ID code \textbf{ovcjhwbhtcpcvwicoxpdpvjzqojril} from \cite{data_arch_SG}, used in Figs. \ref{fig:2stage_sk_qw}, \ref{fig:2stage_bias_sk} and \ref{fig:preanneal_multi}.
  Bottom: the $n=9$ instance from \cite{data_arch_SG} (ID code \textbf{cpahzppaxangdnisyqutdbbjlkqamc}) with the smallest minimum gap.
  \label{fig:SKexample_gapdyn_sparam_both}
  }
\end{figure}

The minimum gap between the ground state and first excited state of the total Hamiltonian is often used in the adiabatic limit of quantum annealing as an indication of the computational difficulty of different parts of the anneal.
Although inspired by the analytical solution to the search problem, where balancing the driver and problem Hamiltonians corresponds to this minimum gap, we have no reason to expect the local Hamiltonians balanced by maximizing $\overline{\mathrm{Dyn}}$ to also locate the global minimum gap, except in special cases.
Figure \ref{fig:SKexample_gapdyn_sparam_both} shows a comparison, for two different $n=9$ SK instances, between the average dynamic coefficient $\overline{\mathrm{Dyn}}$ (red solid line, right axis) and the gap between the ground- and first excited-state (blue solid line, left axis).
The quantities are plotted against the schedule parameter $s(t)$ in the AQC-like paramaterization $A(t)=1-s(t), B(t)=s(t)$.
The maximum of $\overline{\mathrm{Dyn}}$ and the minimum gap are indicated by the red and blue dotted lines respectively.
As in these examples, it is typical, for the SK spin glasses, for the maximum $\overline{\mathrm{Dyn}}$ value to appear significantly before the minimum gap (i.e., closer to the driver end of the schedule).
This could be related to the fact that the smallest gaps occur in a spin-glass phase in which dynamics are expected to be much slower, as described in \cite{knysh2016zero} and discussed in relation to the SK problem in \cite{callison19a}.
Transitioning slowly through the minimum gap is important for the long timescales of adiabatic quantum computing, but it is not necessarily related to what is needed for maximizing the success probability for shorter run times.
Away from the adiabatic limit, there are different mechanisms at play, as has been highlighted by Wong and Meyer \cite{wong16a} and discussed elsewhere \cite{Morley19a,callison19a}.

\subsection{Analytical bounds on \texorpdfstring{$\overline{\mathrm{Dyn}}$}{the average dynamic coefficient}}\label{ssec:AnalBoundDyn}%%%%%%%%%%%%%%%%%%

Equipped with the definition of the average dynamic coefficient $\overline{\mathrm{Dyn}}$, we can investigate when it is possible to find a value of $\Gamma(t)$ such that $\overline{\mathrm{Dyn}}$ is large enough for significant short time dynamics to be generated.
For simplicity, we restrict ourselves to the unbiased driver case,
when the local driver coefficient $D^{(jk)}$ and local transfer coefficient $T^{(jk)}$ are related by $D^{(jk)}=1-T^{(jk)}$.
In this case, the local dynamic coefficient $\mathrm{Dyn}^{(jk)}$ can be written in terms of the driver strength $\Gamma(t)$ and a single \textit{scaled gap} parameter $\zeta_{jk}=\frac{|\Delta_{jk}|}{2|\sandwich{k}{H_\mathrm{drive}}{j}|}$ as
\begin{eqnarray}
\mathrm{Dyn}^{(jk)} & = & \frac{\nicefrac{\zeta_{jk}}{\Gamma(t)}}{(1+\nicefrac{\zeta_{jk}}{\Gamma(t)})^{2}}.\label{eqn:dynjk_unbiased}
\end{eqnarray}
If we write $p_\zeta$ for the probability density function that governs the distribution of $\zeta_{jk}$ in the particular problem and driver Hamiltonians under consideration, then it can be shown that the maximum value attained by the average dynamic coefficient $\overline{\mathrm{Dyn}}$ for any choice of driver strength $\Gamma(t)$ has a lower bound which can be stated formally as
\begin{eqnarray}
& \max_\Gamma(t)(\overline{\mathrm{Dyn}}) \geq & \nonumber\\
& \max_{0<c<1}\left[\frac{1-c}{(2-c)^{2}}\left(1-\frac{1}{c^{2}}\frac{\mu_{2}(p_{\zeta})}{\mu_{1}^{2}(p_{\zeta})}\right)\right], &\label{eqn:dynbar_simplebound}
\end{eqnarray}
where $\mu_1(p_\zeta)$ ($\mu_2(p_\zeta)$) is the first (second) central moment of the distribution governed by the probability density function $p_\zeta$.
Note that this bound is obtained by choosing the specific driver strength $\Gamma=\mu_1(p_\zeta)$, i.e., the mean of the rescaled local gaps, which is not necessarily optimal, but serves to produce a non-trivial lower bound.
We give the proof of the formal lower bound (\ref{eqn:dynbar_simplebound}) in Appendix \ref{app:dynbar_simplebound}.

It is illuminating to look at the shape of this bound, which can be easily computed numerically for any given value of the ratio of moments $\nicefrac{\mu_{2}(p_{\zeta})}{\mu_{1}^{2}(p_{\zeta})}$.
\begin{figure}
\begin{centering}
\includegraphics[width=0.99\columnwidth]{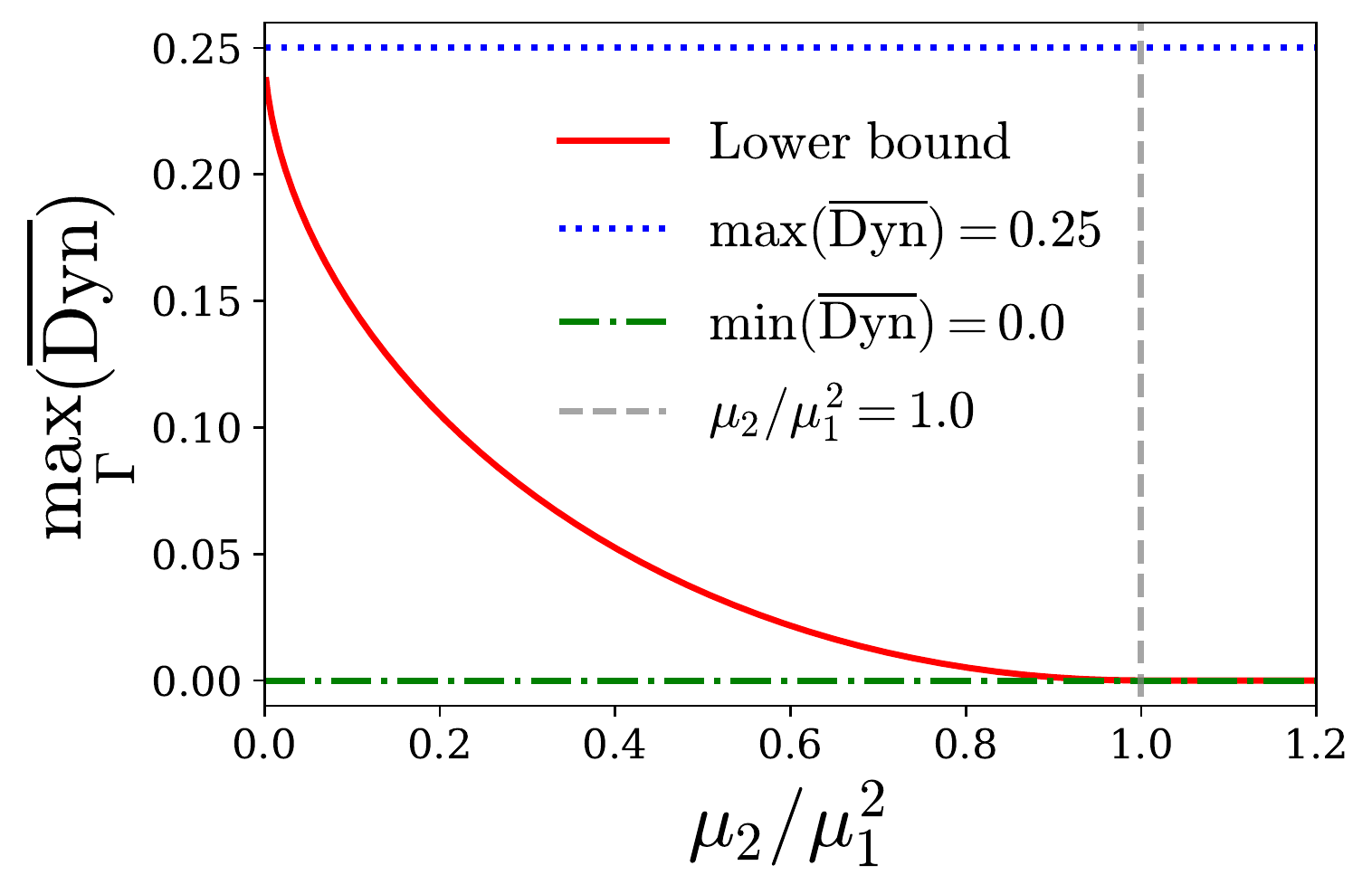} 
\caption{\label{fig:dyn_simplebound}Semi-analytical lower bound (solid, red) on $\overline{\mathrm{Dyn}}$ as a function of ratio of moments $\nicefrac{\mu_{2}(p_{\zeta})}{\mu_{1}^{2}(p_{\zeta})}$ of the distribution, governed by $p_\zeta$, of the rescaled energy gaps $\zeta_{jk}$ between computational basis states connected by the driver Hamiltonian $H_\mathrm{drive}$. 
Also shown, minimum ($0.0$, dot-dashed, green) and maximum ($0.25$, dotted, blue) possible values of $\overline{\mathrm{Dyn}}$.
The lower-bound is non-trivial for  $\nicefrac{\mu_{2}(p_{\zeta})}{\mu_{1}^{2}(p_{\zeta})}<1.0$ (left of grey dashed line), trivially zero otherwise.}
\end{centering}
\end{figure}
The bound is plotted for the interesting range of the ratio of moments in Fig.~\ref{fig:dyn_simplebound}.
It can be seen that the lower-bound is non-trivial when $\nicefrac{\mu_{2}(p_{\zeta})}{\mu_{1}^{2}(p_{\zeta})}<1.0$, 
but is trivially zero otherwise.  This shows that there is a continuous range where $\overline{\mathrm{Dyn}}$ is bounded away from zero, and hence dynamics will definitely happen on short timescales, even for non-optimal choices of $\Gamma(t)$.
This bound is in general far from tight, but still allows us to produce some interesting examples.
We next illustrate the calculation of $\overline{\mathrm{Dyn}}$ and the lower bound in (\ref{eqn:dynbar_simplebound}) for some specific cases.

\subsection{Example: two qubit system}\label{ssec:2qubitDyn}%%%%%%%%%%%%%%%%%%

As a simple example, consider the problem Hamiltonian
\begin{equation}
H_{\mathrm{prob}}^{(2Q)}=-\hat{Z}_1\hat{Z}_2-\frac{1}{2} \hat{Z}_1\nonumber
\end{equation}
as defined in (\ref{eq:H_simple}), with a transverse field driver as defined in (\ref{eqn:transverse_field}). 
For this problem Hamiltonian, there are four two level subspaces connected by the driver, $\ket{00}\leftrightarrow \ket{10}$, $
\ket{00}\leftrightarrow \ket{01}$, $\ket{10}\leftrightarrow \ket{11}$, and $\ket{01}\leftrightarrow \ket{11}$, with $|\Delta_{jk}|=3,2,2,1$ respectively
We can thus calculate $\overline{\mathrm{Dyn}}$ exactly,
\begin{eqnarray}\label{eq:Dyn_bar_simple}
    \overline{\mathrm{Dyn}} & = & \frac{1}{4}\Bigg(\frac{3/(2\,\Gamma)}{(1+3/(2\,\Gamma))^2}+\frac{1/(2\,\Gamma)}{(1+1/(2\,\Gamma))^2} + \nonumber \\
    & &\mbox{\ \ \ \ \ }2\frac{2/(2\,\Gamma)}{(1+2/(2\,\Gamma))^2}\Bigg)\nonumber\\ 
    & = & \frac{\Gamma}{2}\Bigg(\frac{3}{(3+2\,\Gamma)^2}+\frac{1}{(1+2\,\Gamma)^2}+\nonumber \\
    & & \mbox{\ \ \ \ \ }\frac{4}{(2+2\,\Gamma)^2}\Bigg),
\end{eqnarray}
where the time dependence in $\Gamma(t)$ has been omitted for clarity.
To obtain the maximum value of $\overline{\mathrm{Dyn}}$ we need to maximize the expression in (\ref{eq:Dyn_bar_simple}) with respect to $\Gamma$.  
This is easiest done numerically, giving $\max_\Gamma(\overline{\mathrm{Dyn}})\approx 0.241$ for $\Gamma \approx 0.941$. Comparing with the bound in (\ref{eqn:dynbar_simplebound}),
the first moment of $p_\zeta$ is $\mu_1(p_\zeta)=1$, while the second moment is $\mu_2(p_\zeta)=0.125$. 
Based on the ratio $\frac{\mu_2(p_\zeta)}{\mu^2_1(p_\zeta)}=0.125$, we obtain the lower bound $\max_\Gamma \overline{\mathrm{Dyn}}\gtrsim 0.135$. 
This is just over half the actual value, but holds for any Hamiltonian with the same moments of the distribution.

\subsection{Example: Sherrington-Kirkpatrick spin-glass}\label{ssec:SKexample}%%%%%%%%%%%%%%%%%%%%%%%%%%%%%%%%%%%%%

We consider the Sherrington-Kirkpatrick spin-glass problem Hamiltonian given in (\ref{eqn:SKh}). We take the driver Hamiltonian $H_\mathrm{drive}$ to be the transverse field defined in (\ref{eqn:transverse_field}).
Due to the promising results found in \cite{callison19a} for solving this problem with quantum walks, as well as for the more general quenches presented in section \ref{sec:mon_examples}, we expect intuitively that it should be generally possible to find values of $\Gamma$ for which the average dynamic coefficient $\overline{\mathrm{Dyn}}$ takes an appreciable value.

The transverse field driver only connects pairs of states $j,k$ that differ by a single bit flip.
Thus, it can be seen from (\ref{eqn:SKh}) that, for all such pairs, the energy difference can be written
\begin{eqnarray}
\Delta_{jk} &=&  -\sum_{b\neq a}s^{(j)}_{ab} J_{ab} - 2s^{(j)}_a h_a \label{eqn:spinflip_gap}
\end{eqnarray}
where $a$ is the index of the spin that is flipped between states $\ket{j}$ and $\ket{k}$, the sum runs over $b$ which indexes the other spins, $s^{(j)}_{ab}$ is the eigenvalue ($\pm1$) of the operator $\hat{Z}_a \hat{Z}_b$ on the state $\ket{j}$ and $s^{(j)}_{a}$ is the eigenvalue ($\pm1$) of the operator $\hat{Z}_a$ on the state $\ket{j}$.
The gap $\Delta_{jk}$ in (\ref{eqn:spinflip_gap}) is a sum of normally distributed variables with mean $0$, and so $\Delta_{jk}$ is itself a normally distributed variable with mean $0$, and can be shown to have a standard deviation $\varsigma=\sqrt{2(n+1)}\sigma$, where $n$ is the number of spins (qubits).
Then, since $\sandwich{k}{H_\mathrm{drive}}{j}=1$ for the unbiased transverse field driver, the scaled gap $\zeta_{jk}$ is distributed according to the \textit{half-normal} distribution with probability density function
\begin{equation}
p_\zeta(\zeta) = \frac{1}{\varsigma\sqrt{2\pi}}\exp\left(-\frac{\zeta^2}{8\varsigma^2}\right),\,\,\,\zeta \geq 0 \label{eqn:halfnormal}
\end{equation}
For this distribution, it can be shown that the ratio of moments is
\begin{eqnarray}
\frac{\mu_2(p_\zeta)}{\mu^2_1(p_\zeta)} &=& \frac{1-\nicefrac{2}{\pi}}{\nicefrac{2}{\pi}} \nonumber\\
& \approx & 0.571,
\end{eqnarray}
which we emphasise is independent of the width $\varsigma$ of the distribution of the scaled gap $\zeta_{jk}$.
For this value of the ratio, the lower bound shown in Fig. \ref{fig:dyn_simplebound} is
\begin{eqnarray}
\max_\Gamma (\overline{\mathrm{Dyn}}) & \gtrsim & 0.03.
\end{eqnarray}
While this value is small compared to the maximum possible value of $\overline{\mathrm{Dyn}}=0.25$, which is not unexpected for a hard problem (NP hard), we emphasise that it is independent of the width $\varsigma$ of the distribution of the scaled gap $\zeta_{jk}$ and thus does not scale with the system size.  
Bounding $\overline{\mathrm{Dyn}}$ away from zero for all sizes proves that dynamics will occur over short timescales for suitable control parameters, thus providing evidence that the scaling found in \cite{callison19a} may continue to useful problem sizes.

\subsection{Example: unstructured search\label{ssec:search_dyn_bound}} %%%%%

As a contrasting example, we consider the problem of unstructured search on $n$ qubits, in which a single computational basis state  $\ket{m}$, out of the total $N=2^n$ basis states, is marked by being given a lower energy.
The Hamiltonian for this problem is
\begin{equation}
H_{\mathrm{search}}=\openone-2\ketbra{m}{m}.
\end{equation}
and again we take the driver Hamiltonian $H_\mathrm{drive}$ to be the transverse field defined in (\ref{eqn:transverse_field}).
While unstructured search is a well known example with a provable quantum advantage, the algorithms which yield this advantage all involve coherent operations on time scales of order $\sqrt{N}=2^{\frac{n}{2}}$ rather than the short-time dynamics we are discussing in this paper. As such, we would intuitively not expect the lower bound in (\ref{eqn:dynbar_simplebound}) to be large in this case.

Of the $n\,2^{n-1}$ total off diagonal matrix element pairs in the transverse field driver, only $n$ of these will connect a pair of computational basis states with non-zero energy difference, having energy difference $\Delta_{jk} = 2$, with the remaining $n\,2^{n-1}-n$ pairs having zero-energy difference  $\Delta_{jk} = 0$. 
Therefore, the distribution of scaled gaps $\zeta_{jk}$ can be written as 
\begin{eqnarray}
p_\zeta (\zeta) & = & \frac{n}{n\,2^{n-1}}\delta(\zeta-1) + \left(1-\frac{n}{n\,2^{n-1}}\right)\delta(\zeta)
\end{eqnarray}
Calculating the first and second central moments of this distribution gives
\begin{eqnarray}
\mu_1(p_\zeta) &=& \frac{1}{2^{n-1}} \\
\mu_2(p_\zeta) &=& \frac{1}{2^{n-1}} - \left(\frac{1}{2^{n-1}}\right)^2
\end{eqnarray}
and so the relevant ratio of moments is
\begin{eqnarray}
\frac{\mu_2(p_\zeta)}{\mu^2_1(p_\zeta)} & = & 2^{n-1} - 1
\end{eqnarray}
Looking at the plot of the lower bound in Fig. \ref{fig:dyn_simplebound}, we can see that, for unstructured search the bound is trivially zero for all $n>1$ 
We can also calculate the exact value using (\ref{eqn:dynjk_unbiased}).  
For each of the $n\,2^{n-1}-n$ pairs of states $j,k$ with $|\Delta_{jk}|=0$, $\mathrm{Dyn}^{(jk)}=0\, \forall \Gamma$.
For the remaining $n$ pairs of states $j,k$ with $|\Delta_{jk}|=2$, the choice of driver strength $\Gamma=1.0$ will maximise $\mathrm{Dyn}^{(jk)}=0.25$.
Thus, the average dynamic coefficient for unstructured search is
\begin{eqnarray}
\overline{\mathrm{Dyn}}&=&\frac{1}{2^{n-1}}\times0.25 \nonumber\\
 &=& \frac{1}{2^{n+1}} \label{eq:search_dyn}
\end{eqnarray}
which tends toward the lower bound of zero in the limit as $n\rightarrow\infty$.

This tells us that, for search, most two-level subspaces do not exhibit dynamics and probability enhancement of the marked state can only happen through finely tuned control. 
For an adiabatic algorithm, this is achieved by slowly adjusting the Hamiltonian within a precise range so that the system can follow a very delicate path, whereas for quantum walk this is achieved by reaching a finely tuned resonance between the marked state and the rest of a symmetric subspace of the Hilbert space. While interpolations between these two extremes are possible \citep{Morley19a}, all of the interpolated algorithms also rely on dynamics of a two level system with a gap proportional to $\sqrt{N}=2^{\frac{n}{2}}$. 
In such a system, significant dynamics cannot occur in the timescales of rapid quenches, $O(1)$ or $O(\mathrm{poly}(n))$.

%%%%%%%%%%%%%%%%%%%%%%%%%%%%%%%%%%%%%%%%%%%%%%%%
\section{Using dynamics to find heuristic quench parameters\label{sec:s_heur}}

As mentioned in section \ref{sec:ensure_dyn}, the average dynamic coefficient $\overline{\mathrm{Dyn}}$ can in general be efficiently estimated by sampling.
In this section, we show via two practical examples that this estimate can be used to develop heuristic methods for setting the control function $\Gamma(t)$, or equivalently, $A(t)$ and $B(t)$, for a rapid quench, in both quantum walk and quantum annealing settings.
In both cases, we use the unbiased transverse field driver Hamiltonian defined in (\ref{eqn:transverse_field}).
First, we consider the quantum walk algorithm, starting with a simplified example of a two qubit system.  We then develop a heuristic for the Sherrington-Kirkpatrick spin-glass, and show that it performs almost as well as the numerically fine-tuned heuristic described in \cite{callison19a}, without needing any fine-tuning. 
Second, we develop a simple heuristic method for defining a schedule for a time-dependent rapid quench, also applied to the Sherrington-Kirkpatrick spin-glass, that outperforms a linear ramp.

In all the examples discussed in this section, we computed the average dynamic coefficient $\overline{\mathrm{Dyn}}$ numerically using all non-zero $j$,$k$ pairs, rather than estimating it by sampling such pairs.
This is computationally easy to do at these problem sizes, and allows us to separate the effectiveness of the heuristic from errors due to sampling.

\subsection{Heuristic hopping rate for a quantum walk\label{ssec:SK_dyn_gamma}}%%%%%%%%%%%%%%%%%%%%%%%%

For a quantum walk, the average dynamic coefficient $\overline{\mathrm{Dyn}}$ is a function of the chosen hopping rate $\Gamma(t)=\gamma$.
Informed by the result in section \ref{sec:en_perspect} that dynamics will typically be useful, 
it follows that by maximizing $\overline{\mathrm{Dyn}}$ we can obtain a 
heuristic hopping rate $\gamma_\mathrm{Dyn}$,
that should ensure significant dynamics occur over short timescales.
\begin{figure}
\begin{centering}
\includegraphics[width=0.99\columnwidth]{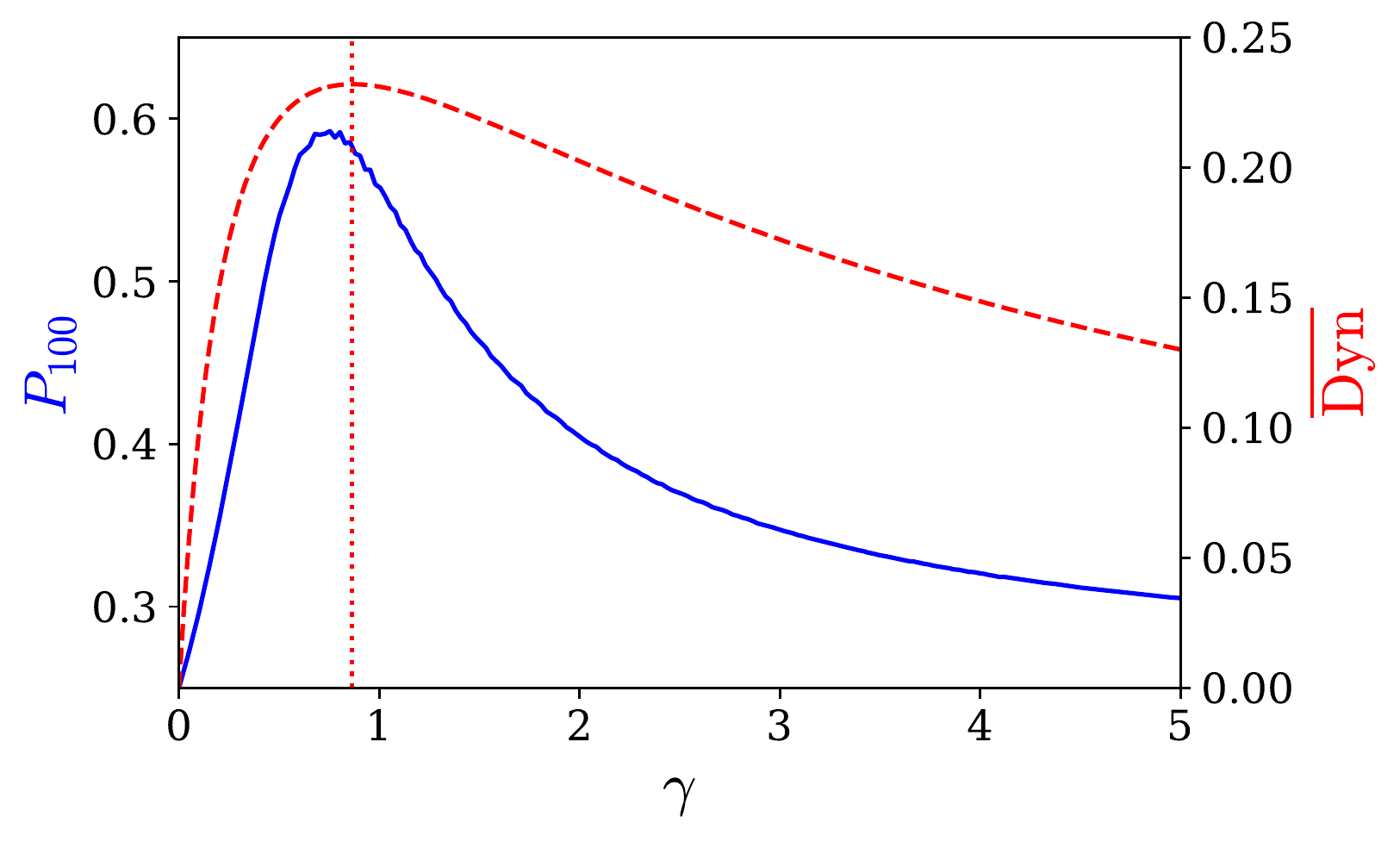}
\caption{\label{fig:2qb_Dynbar_lta} 
Average success probability $p_{100}$ between $t=0$ and $t=100$ (blue, solid), calculated based on $10,000$ independent random points within this range and $\overline{\mathrm{Dyn}}$ (red, dashed) versus $\gamma$ for the two qubit system given in (\ref{eq:H_simple}). Dotted vertical line indicates the value of $\gamma_{\mathrm{Dyn}}$.
}
\end{centering}
\end{figure}
For the two qubit Hamiltonian from (\ref{eq:H_simple}), Fig.~\ref{fig:2qb_Dynbar_lta} shows how the average success probaility within $100$ dimensionless time units $P_{100}$ varies with $\gamma$.  
For this two qubit system, we can exactly calculate $\overline{\mathrm{Dyn}}$, see section \ref{ssec:2qubitDyn}, shown in Fig.~\ref{fig:2qb_Dynbar_lta}.
The maximum value of $\overline{\mathrm{Dyn}}$ gives a value for $\gamma_{\mathrm{Dyn}}$ which is a good quality estimate for the value of $\gamma_{\mathrm{opt}}$.
Using bisection and a numerically calculated derivative, we find that  $\gamma_{\mathrm{Dyn}}\approx 0.864$, while  the peak of $P_{100}$ occurs at a slightly lower value of $\gamma$. 
Since the peak of $P_{100}$ is quite broad, the discrepancy between $\gamma_{\mathrm{Dyn}}$ and $\gamma_{\mathrm{opt}}$ only reduces $P_{100}$ by a small amount, as can be seen in Fig.~\ref{fig:2qb_Dynbar_lta}.  

\begin{figure}
\begin{centering}
\includegraphics[width=0.99\columnwidth]{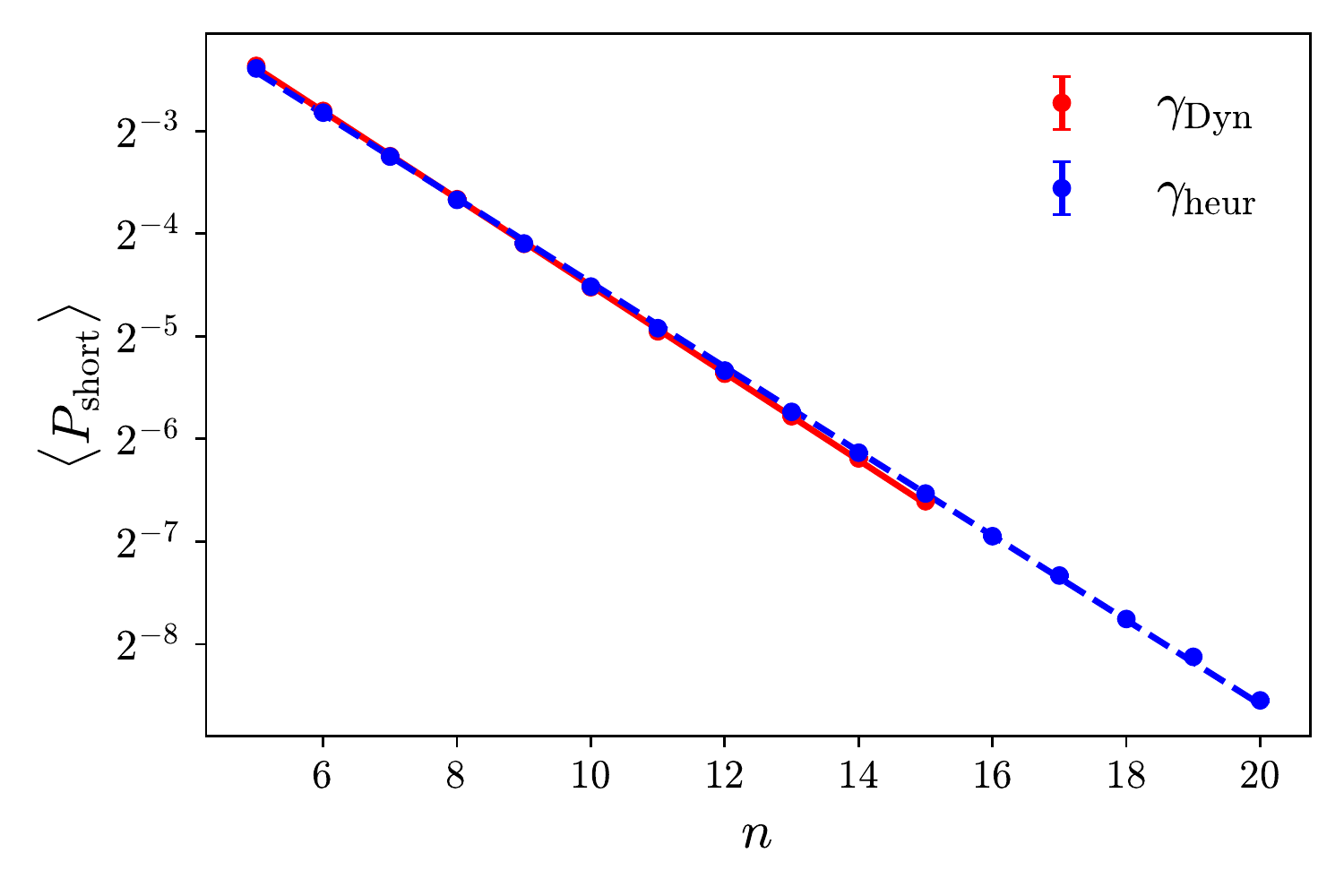}
\caption{\label{fig:gdyn_probscaling} Log-linear plot of average \textit{short time} success probability $\langle P_\mathrm{short}\rangle$ against number of qubits $n$ for quantum walks on the spin-glass dataset from \cite{data_arch_SG}, using the heuristic hopping rate $\gamma_\mathrm{dyn}$ derived for each instance by optimizing the average dynamic coefficient $\overline{\mathrm{Dyn}}$ (red).
Also shown for comparison, $\langle P_\mathrm{short}\rangle$ obtained using the fine-tuned heuristic hopping rate $\gamma_\mathrm{heur}$ (blue) described in \cite{callison19a}.}
\end{centering}
\end{figure}
To test how well this heuristic hopping rate works for a more realistic example, we numerically calculated $\gamma_\mathrm{Dyn}$ 
for each instance of size $5\leq n\leq15$ of the spin glass problems from \cite{data_arch_SG}.
This was done by performing a bisection optimization to maximise the value of $\overline{\mathrm{Dyn}}$ as a function of $\gamma$ for each instance.
Following the methods in \cite{callison19a}, we performed a short-time quantum walk and calculated the success probability $P_\mathrm{short}$, which is time-averaged over a short run time.
Averaging over all instances of a given size, we obtain the average \textit{short time} success probability 
\begin{eqnarray}
 P_\mathrm{short} & = & \intop_{\frac{12.5}{\sqrt{n}}}^{\frac{17.5}{\sqrt{n}}}\mathrm{d}t\, P(t),
\end{eqnarray}
defined in \cite{callison19a}), for measuring the problem ground-state.
This is shown (red line) for each size in Fig.~\ref{fig:gdyn_probscaling}.
Included for comparison (blue line) are the results from \cite{callison19a} using the fine-tuned heuristic $\gamma_\mathrm{heur}$ defined there, using properties of the eigenvalue distribution for the spin glass problem Hamiltonian.
It can be seen that, despite $\gamma_\mathrm{heur}$ being numerically fine-tuned specifically for the Sherrington-Kirkpatrick spin-glass problem, it performs only marginally better the general method we have used here.  Fitting the data produces $\langle P_\mathrm{short}\rangle \sim O(N^{(-0.411\pm0.002)})$ for $\gamma_\mathrm{heur}$ compared to $\langle P_\mathrm{short}\rangle \sim O(N^{(-0.425\pm0.001)})$ for $\gamma_\mathrm{Dyn}$.
The eigenvalue distribution used in \cite{callison19a} would not generally be available to calculate $\gamma$ for real problems; this comparison shows that using $\overline{\mathrm{Dyn}}$ is a viable method for determining a useful value for $\gamma$ in this case.

For the small size instances we are using, we have used all the values of $\mathrm{Dyn}^{(jk)}$ to calculate the average in the definition of $\overline{\mathrm{Dyn}}$ in (\ref{eq:dynbar_def}).  We can show that the error in $\gamma_{\mathrm{Dyn}}$ due to sampling a subset of $\mathrm{Dyn}^{(jk)}$ values stays manageable for larger sizes. 
Consider a small error $\delta\gamma$ in $\gamma$.  
Doing a Taylor expansion of  $\overline{\mathrm{Dyn}}(\gamma)$ around its peak value $\overline{\mathrm{Dyn}}_{\mathrm{max}}$ gives
\begin{eqnarray}
    & \delta \overline{\mathrm{Dyn}} =  \overline{\mathrm{Dyn}}_{\mathrm{max}} - \overline{\mathrm{Dyn}}(\gamma_{\mathrm{Dyn}}+\delta \gamma) & \nonumber \\
    & =  -(\delta \gamma)^2\left.\frac{\partial^2 \overline{\mathrm{Dyn}}(\gamma)}{\partial \gamma^2}\right|_{\gamma=\gamma_{\mathrm{Dyn}}}+O\left((\delta \gamma)^3\right), & \label{eq:Dyn_bar_expand}
\end{eqnarray}
where $\gamma_{\mathrm{Dyn}}$ is the value of $\gamma$ our heuristic would find %
\footnote{Note that the first order term is absent, and the second order term is negative, because we are sampling at the maximum, and we have implicitly assumed that there are no other peaks in $\overline{\mathrm{Dyn}}$ which take similar values.  If $\left.\frac{\partial^2 \overline{\mathrm{Dyn}}(\gamma)}{\partial \gamma^2}\right|_{\gamma=\gamma_{\mathrm{Dyn}}}$ vanishes, then the next non-zero even derivative should be used.}
with the exact $\overline{\mathrm{Dyn}}_{\mathrm{max}}$.
Using the sampling error in $\overline{\mathrm{Dyn}}$ from (\ref{eq:deltaDynbar})
and rearranging yields
\begin{equation}\label{eq:delta_gamma}
\delta \gamma\propto N^{-\frac{1}{4}}_{\mathrm{sample}}\left(-\left.\frac{\partial^2 \overline{\mathrm{Dyn}}(\gamma)}{\partial \gamma^2}\right|_{\gamma=\gamma_{\mathrm{Dyn}}}\right)^{-\frac{1}{2}}.
\end{equation}
This is a general expression that can be used for any problem Hamiltonian.
For the Sherrington-Kirkpatrick spin glass, we can use the distribution of the scaled
gaps from (\ref{eqn:halfnormal}),
and the definition of $\mathrm{Dyn}^{(j,k)}$ from (\ref{eqn:dynjk_unbiased}),
to obtain the average value of $\overline{\mathrm{Dyn}}(\gamma)$ for SK instances, $\langle\overline{\mathrm{Dyn}}\rangle (\gamma)$
\begin{eqnarray}
& \langle\overline{\mathrm{Dyn}}\rangle(\gamma)=& \nonumber \\ &\frac{1}{\varsigma\sqrt{2\pi}}\int_0^\infty d\zeta \exp\left\{-\frac{\zeta^2}{8\varsigma^2}\right\}\frac{\nicefrac{\zeta}{\gamma}}{(1+\nicefrac{\zeta}{\gamma})^{2}}.& \label{eq:Dynbarav}
\end{eqnarray}
Making the substitution $z=\nicefrac{\zeta}{(2\sqrt{2}\varsigma)}$, to remove the $\varsigma$ dependence in the exponential, and differentiating twice w.r.t.~$\gamma$ gives
\begin{eqnarray}
& \frac{\partial^2}{\partial\gamma^2}\langle\overline{\mathrm{Dyn}}\rangle(\gamma)= & \nonumber\\
&\frac{8}{\varsigma^2\sqrt{2\pi}}\int_0^\infty dz\,2z\, \exp\left\{-z^2\right\}
\frac{(\gamma/\varsigma)-4\sqrt{2}z}{\{(\gamma/\varsigma)+2\sqrt{2}z\}^{4}}.& \label{eq:doublederiv}
\end{eqnarray}
This needs to be evaluated at $\gamma=\gamma_{\mathrm{Dyn}}$, at the peak of $\langle\overline{\mathrm{Dyn}}\rangle(\gamma)$, which doing the substitution $z=\nicefrac{\zeta}{(2\sqrt{2}\varsigma)}$ in (\ref{eq:Dynbarav}) shows occurs at a fixed value of $\nicefrac{\gamma}{\varsigma}$.  Hence, the scaling with $n$ of the double derivative at $\gamma=\gamma_{\mathrm{Dyn}}$ is determined solely by the $\varsigma^{-2}$ prefactor in (\ref{eq:doublederiv}).
Recalling from section \ref{ssec:SKexample} that $\varsigma = \sqrt{2(n+1)}$ for these SK spin glasses, and putting it back into (\ref{eq:delta_gamma}) we have
\begin{equation}\label{eq:delta_gamma_SK}
\delta \gamma\propto N^{-\frac{1}{4}}_{\mathrm{sample}}(n+1)^{1/2}.
\end{equation}
The peak in the success probability as a function of $\gamma$ is very broad for SK spin glasses, and the width of this peak decreases as $\nicefrac{1}{n}$ (determined numerically \cite{callison19a}).  
Combined with (\ref{eq:delta_gamma_SK}), this means the sampling rate to calculate $\overline{\mathrm{Dyn}}$ needs to increase by a $\mathrm{poly}(n)$ factor as $n$ increases, in order to determine $\gamma_{\mathrm{Dyn}}$ to sufficient accuracy.  Since $n$ corresponds to the number of qubits, this can be done efficiently.

\subsection{Heuristic schedule for quantum annealing\label{ssec:qa_dyn_heur}}%%%%%%%%%%%%%%%%%%%%%%%%%%%%%%%%%%

\begin{figure}
  \subfigure[]{\includegraphics[width=0.99\columnwidth]{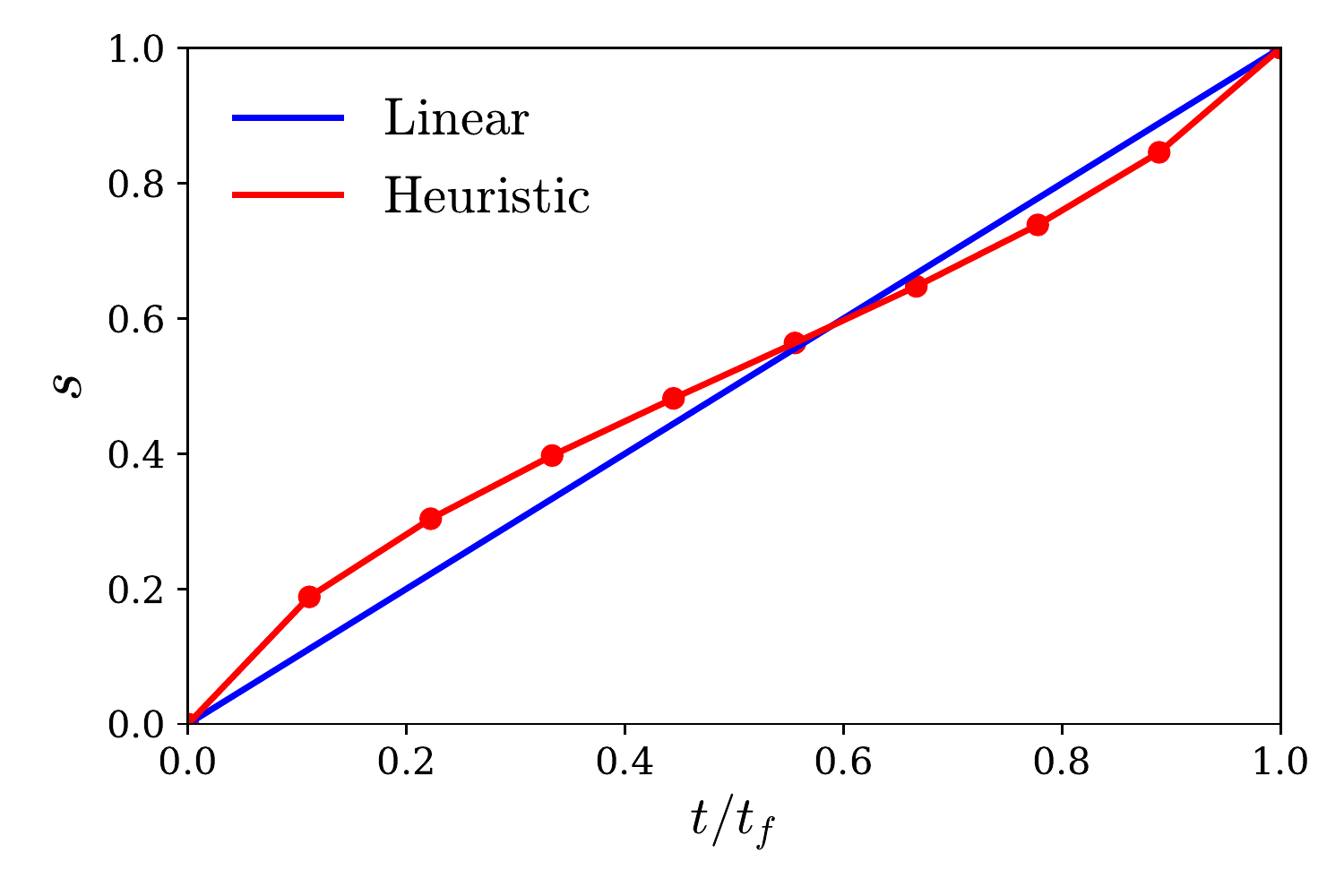}\label{fig:heur_sched}}
  \subfigure[]{\includegraphics[width=0.99\columnwidth]{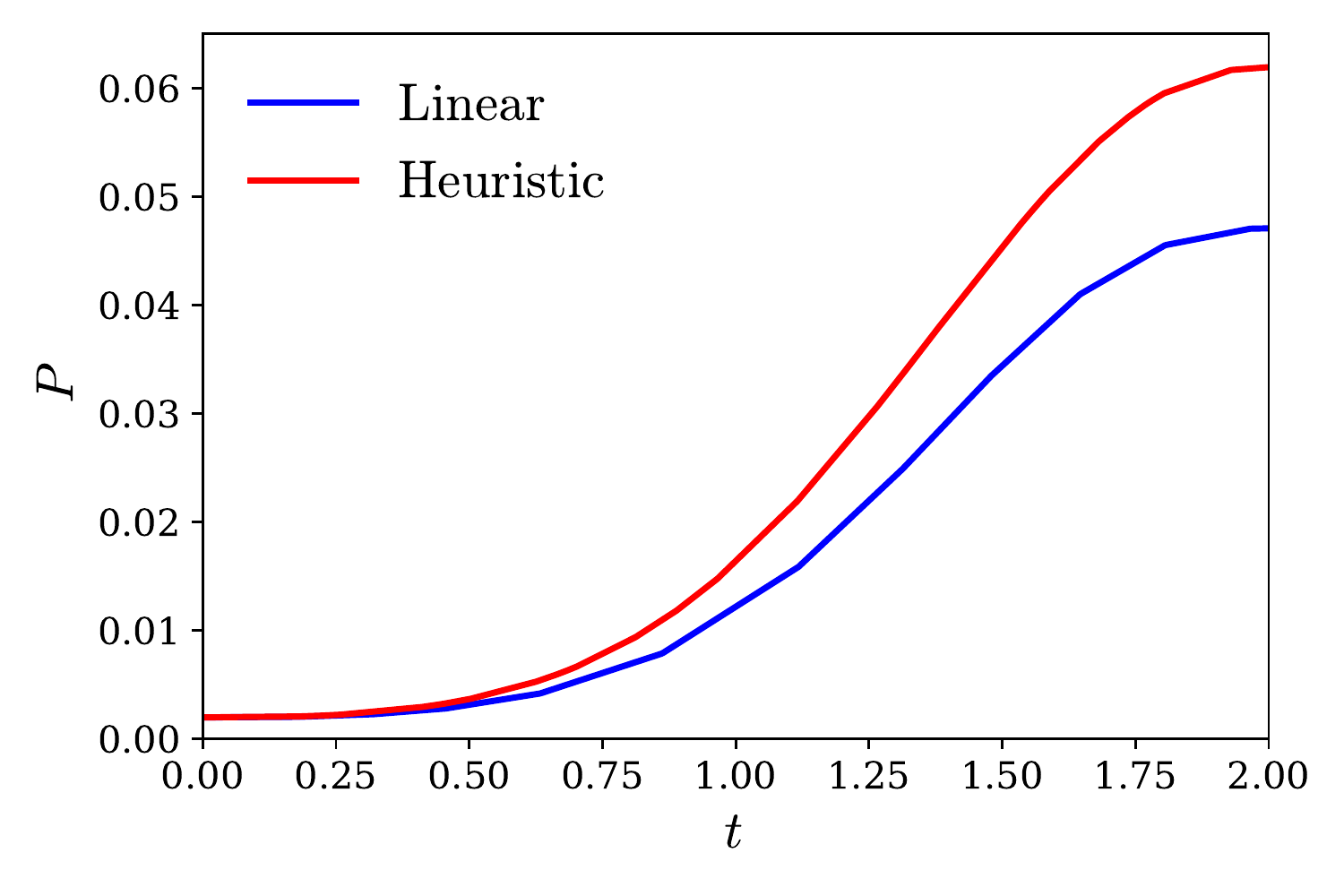}\label{fig:heur_sched_prob}}  
  \caption{Top: A heuristic quench schedule of duration $t_f=2.0$ (red) derived from the average dynamic coefficient $\overline{\mathrm{Dyn}}$ for a typical nine qubit spin-glass instance from \cite{data_arch_SG}. 
  For comparison, a linear schedule (also of duration $t_f=2.0$) is also shown.
  Bottom: The instantaneous success probability $P(t)$ for measuring the problem ground-state for for each time $t$ as the quench progresses along the heuristic schedule (red) and the linear schedule (blue).
  \label{fig:heur_schedule}
  }
\end{figure}

For a time-dependent rapid quench of the form $H_{AB}(t)$ defined in (\ref{eq:Hopt}) and total duration $t_f$, a common choice of control functions, inspired by the adiabatic algorithm, is $A(t)=1-s(t)$ and $B(t)=s(t)$, where $s(t)$ is a schedule function with boundary conditions $s(0)=0$ and $s(t_f)=1$.
In the absence of any knowledge about where along the schedule useful computation can happen, the schedule function is often set to be the linear function $s(t)=\nicefrac{t}{t_f}$.
The average dynamic coefficient $\overline{\mathrm{Dyn}}$ provides a measure of the level of dynamics at each point along the schedule.
Intuition gained from section \ref{sec:en_perspect} suggests that the linear schedule can in general be improved by spending less time in regions where $\overline{\mathrm{Dyn}}$ is small and more time in regions where $\overline{\mathrm{Dyn}}$ is large.
A straightforward way to do this is to choose $\dd{s}{t}\propto\frac{1}{\overline{\mathrm{Dyn}}}$ (the constant of proportionality is set by the boundary conditions $s(0)=0$ and $s(T)=1$). 
We have approximated such a schedule for a typical nine qubit Sherrington-Kirkpatrick spin-glass instance, as shown in Fig.~\ref{fig:heur_sched} (red line).
We have done this by fixing the value of the points marked by circles according to $\Delta s\propto\frac{\Delta t}{\overline{\mathrm{Dyn}}}$, subject to the boundary conditions, and then linearly interpolating between them.
A linear schedule $s(t)=\nicefrac{t}{t_f}$ (blue line) is also shown for comparison.
Figure \ref{fig:heur_sched_prob} shows the instantaneous success probability $P(t)$ for measuring the problem ground-state as the quench progresses along the heuristic schedule (red line) and the linear schedule (blue line) for quench duration of $t_f=2$.
It can be seen that the simple heuristic we've used here has resulted in a significant improvement in success probability at the end of the schedule.  We have checked sufficiently many of the instances to determine that this level of improvement is typical for this size of problem and total time duration $t_f=2$.
Further improvements may be available by varying $t_f$ or choosing a different function of $\overline{\mathrm{Dyn}}$ for $\dd{s}{t}$.

\section{Numerical methods}\label{sec:num_meth} %%%%%%%%%%%%%%
Numerical simulation and optimization were used extensively throughout this work, as much of the analysis we have performed is not analytically tractable.
The simulations and plots were performed using the Python language \citep{van2003python}, aided extensively by the NumPy \citep{oliphant2006guide}, SciPy \citep{scipy}, quimb, \citep{gray2018quimb}, and Matplotlib \citep{hunter2007matplotlib} libraries.
We also used the IPython interpreter \citep{perez2007ipython} and Jupyter notebook system \citep{jupyter}. 
MATLAB was used for some early numerical experiments, but not for any results which directly appear in the manuscript.

The numerical optimization used to produce Figs.~\ref{fig:dyn_simplebound}, \ref{fig:2qb_Dynbar_lta},  \ref{fig:gdyn_probscaling} and \ref{fig:heur_schedule}, as well as the curve fitting used in Figs.~\ref{fig:preanneal_scaling} and \ref{fig:gdyn_probscaling}, was performed using the optimization tools in SciPy \citep{scipy}.

The Sherrington-Kirkpatrick spin glass instances in the data repository at \cite{data_arch_SG} have been used extensively.
In any cases where a single example Sherrington-Kirkpatrick spin-glass instance has been used, it is the instance \textbf{ovcjhwbhtcpcvwicoxpdpvjzqojril}.
The plot of average short time success probability $\langle P_\mathrm{short}\rangle$ against number of spins $n$ in Fig. \ref{fig:gdyn_probscaling} uses all of the Sherrington-Kirkpatrick spin glass instances in the repository.

\section{Summary and further work}\label{sec:conc}

In this paper, we have generalised and extended work begun in \cite{callison19a} to time-varying quantum annealing schedules.
In \cite{callison19a}, Callison et.~al provide numerical evidence for the ability of quantum walks to solve NP hard problems using many repeats of short runs.
This strategy scales better than quantum search, by exploiting the correlations in the problem Hamiltonian.  
The energy conservation mechanism identified in \cite{callison19a} explains how energy conserving quantum walks can find lower energy states with better than guessing probability.
In section \ref{sec:en_perspect}, we generalised the energy conservation mechanism to an \emph{energy redistribution mechanism} that holds for all monotonic quenches which start in the ground state of the driver Hamiltonian and have non-negative control functions. 
This thus includes a wide range of quantum annealing protocols used in both theoretical and experimental work.
The improvements leveraged by time-varying rapid quenches can be considerable, as we illustrated in section \ref{sec:mon_examples}.

To generate significant energy redistribution, there needs to be significant dynamics driving the system away from the initial state.
To characterise the dynamics, in section \ref{sec:ensure_dyn} we defined the average local dynamic coefficient that balances the contributions from both the driver and problem Hamiltonians.
This allows the control functions in the Hamiltonian to be optimised for fast dynamics, and provides a very general way to estimate good values to use for specific problems.
For the spin glass data \citep{chancellor19a}, we showed in Fig.~\ref{fig:gdyn_probscaling} that such estimates are almost as good as the numerically optimised values used in \cite{callison19a}.
We also verified in section \ref{ssec:search_dyn_bound} that our average local dynamic coefficient correctly predicts that the search problem will not have significant dynamics on short timescales.
The average dynamic coefficient we have defined is one way to capture the local dynamics in a quantum annealing Hamiltonian system; doubtless there are other formulations that would serve equally well.  In the transverse Ising setting, it focuses on single spin flips, which intuitively are likely to provide the fastest dynamics.  Settings with driver Hamiltonians applying multiple spin flips (e.g., \cite{roland02a, childs03a}) may prove less favorable for obtaining fast dynamics, a worthwhile direction for future investigations.

Taken together, the \emph{energy redistribution mechanism} and the \emph{average dynamic coefficient} are powerful tools for understanding, designing, and optimally controlling rapid quench quantum annealing algorithms.
We provide a simple example of how to do this to good effect for annealing schedules in section \ref{ssec:qa_dyn_heur}, and verify in section \ref{ssec:SK_dyn_gamma} that it is both efficient and effective for estimating hopping rates for quantum walks on spin glasses.
While adiabatic quantum computing and quantum walk search have long had theoretical underpinnings, this represents a significant step in understanding how to exploit quantum annealing schedules run for short times.
For current state-of-the-art noisy quantum computers, short run times are a big advantage over the long coherence times required for adiabatic quantum computing, or quantum walk search.

We have shown that our tools apply to the biased drivers proposed in  \cite{Duan2013a,Grass19a}, which provide a method of incorporating prior information into annealing schedules.
This can produce significant improvements, as we illustrate in section \ref{ssec:biased2QW}.
On the other hand, reverse annealing schedules, both as proposed by \cite{Perdomo-Ortiz11,Ohkuwa18a,Yamashiro19a} and discussed in \cite{chancellor17b}, and as implemented in the latest D-Wave Systems \citep{reverse_anneling_whitepaper},
are by definition not monotonic, so the tools and mechanisms identified here cannot be applied.
Since reverse annealing is a powerful tool, extending our results to non-monotonic cases is an important direction for further research.

\section*{Acknowledgements}%%%%%%%%%%%%%%%%%%%

We thank Jemma Bennett for providing useful references related to error correction, and David Ross for spotting an error in one of the figures. NC and JC were supported by UKRI EPSRC grant EP/S00114X/1. LN was supported by a Durham University studentship. AC was funded by UKRI EPSRC grant EP/L016524/1 via the Imperial College London CDT in Controlled Quantum Dynamics. VK and NC were supported by UKRI EPSRC grant EP/L022303/1. 

%%%%%%%%%%%%%%%%%%%%%%%%%%%%%%%%%%%%%%%%%%%%%%
\appendix

\section{Proof: monotonic quenches do no worse than guessing \label{app:energy_redist_proof}}

\subsection{Energy conservation mechanism}\label{app:enconsmech}

In this appendix, we recap the special case presented in \cite{callison19a,Hastings19a} for time independent controls.   Quantum walks can be viewed as a closed-system annealing protocol with a discontinuous schedule \citep{Morley19a}. For QW, when formulated in terms of Eq.~(\ref{eq:Hopt}) $A(t)$ and $B(t)$ are constant, independent of time. This picture however doesn't follow the convention of how annealing protocols are formulated, where the system starts in the ground state of the initial Hamiltonian and the driver is completely absent at the end of the anneal. Following such a convention is important for instance to define an interpolation between annealing protocols and QW, as was done in \citep{Morley19a}.  To define QW as an annealing protocol in which $A(0)=B(t_{f})=1$ and $A(t_{f})=B(0)=0$, we can write $A(t)=\gamma\,\Theta(t_{f}-t+\epsilon)$ and $B(t)=\Theta(t-\epsilon)$, where $\Theta$ is the Heaviside theta function, $\Theta(a>0)=1$, $\Theta(a<0)=0$, $\Theta(a=0)=\frac{1}{2}$,and take the limit where $\epsilon\rightarrow 0$. 

Since the initial state $\ket{\psi(t=0)}$ is a ground state of the driver Hamiltonian $H_{\mathrm{drive}}$, it follows immediately that the expectation value of the driver Hamiltonian is at its lowest at $t=0$, that is, $\sandwich{\psi(t)}{H_\mathrm{drive}}{\psi(t)} \ge \sandwich{\psi(t=0)}{H_\mathrm{drive}}{\psi(t=0)}$, since the expectation value of the driver Hamiltonian $H_{\mathrm{drive}}$ for any quantum state cannot be less than that of the ground state.

The total energy expectation as a function of time can be written
\begin{eqnarray}
E(t)&=&\sandwich{\psi(t)}{\gamma H_{\mathrm{drive}}+H_{\mathrm{prob}}}{\psi(t)} \nonumber \\
&=&\gamma \average{H_{\mathrm{drive}}}_{\psi(t)}+\average{H_{\mathrm{prob}}}_{\psi(t)},
\end{eqnarray}
where the notation $\langle . \rangle_{\psi}$ is used to denote the expectation value with respect to the state $| \psi \rangle$.
Since energy is conserved for $0<t<t_f$, it follows that, for $\epsilon\rightarrow 0$, $E(\epsilon)=E(t_f-\epsilon)$, and therefore
\begin{eqnarray}
&\gamma \average{H_{\mathrm{drive}}}_{\psi(t=0)}+\average{H_{\mathrm{prob}}}_{\psi(t=0)}=&\nonumber\\ 
&\gamma \average{H_{\mathrm{drive}}}_{\psi(t_f)}+\average{H_{\mathrm{prob}}}_{\psi(t_f)}&
\end{eqnarray}
rearranging terms, and recalling that $\psi(t=0)$ is the ground state of $H_{\mathrm{drive}}$ and $\gamma \ge 0$, we observe that,
\begin{eqnarray}
&\average{H_{\mathrm{prob}}}_{\psi(t_f)}-\average{H_{\mathrm{prob}}}_{\psi(t=0)}=&\nonumber\\
&\gamma[ \average{H_{\mathrm{drive}}}_{\psi(t=0)}-\average{H_{\mathrm{drive}}}_{\psi(t_f)}]\le 0 &,
\end{eqnarray}
and therefore $\average{H_{\mathrm{prob}}}_{\psi(t_{\mathrm{f}})} \le \average{H_{\mathrm{prob}}}_{\psi(t=0)}$.
Since $\psi(t=0)$ is not an eigenstate of the full Hamiltonian, some dynamics are guaranteed to happen, and thus there will be times $t>0$ when $\average{H_{\mathrm{prob}}}_{\psi(t)}$ is strictly less than $\average{H_{\mathrm{prob}}}_{\psi(t=0)}$.
%In the next subsection we show that by discretizing and using a rescaled representation, we can use a combination of the energy conservation mechanism and arguments about the stages where the Hamiltonian is modified, to generalize a version of the energy redistribution mechanism to all cases of time dependent Hamiltonian evolution which obey the conditions given in section \ref{sec:en_perspect}.

\subsection{\texorpdfstring{Energy redistribution mechanism in the case of $B(t)\rightarrow 0$: divergence of $\Gamma$}{The case of B(t)=0: divergence of Gamma}}\label{ssec:divergence_of_gamma}
The result in section \ref{sec:en_perspect} is that the inequality (\ref{eqn:quench_beats_guessing}) holds for any quench with a Hamiltonian in the form of (\ref{eq:Hopt_Gamma})
that satisfies the three conditions listed in section \ref{sec:en_perspect}.
We now consider quenches with a Hamiltonian in the form of (\ref{eq:Hopt}).
Any Hamiltonian of the form (\ref{eq:Hopt}) with $B(0)>0$ can be put in the form of (\ref{eq:Hopt_Gamma}) by identifying the ratio $\nicefrac{A(t)}{B(t)}$ with $\Gamma(t)$ and rescaling by a factor $\nicefrac{1}{B(t)}$, which can be formally compensated for by rescaling time by a factor of ${B(t)}$.
Thus, the inequality (\ref{eqn:quench_beats_guessing}) holds also for any quench with a Hamiltonian in the form of (\ref{eq:Hopt}) with $B(0)>0$ and which otherwise satisfies the three conditions listed in section \ref{sec:en_perspect}.
Here, we show that this can be extended to to case where $B(0)=0$.

In the case that $B(0)=0$, consider the modified Hamiltonians
\begin{eqnarray}
H'_\mathrm{drive} &=& H_\mathrm{drive} - \frac{\epsilon}{A(0)}H_\mathrm{prob} \label{eqn:mod_driver}\\
H'_\mathrm{prob} &=& H_\mathrm{prob}
\end{eqnarray}
and the modified control functions
\begin{eqnarray}
A'(t) &=& A(t) \\
B'(t) &=& B(t) + \frac{A(t)}{A(0)}\epsilon \nonumber \\
&=& B(t)\left[1+\Gamma(t)\frac{\epsilon}{A(0)}\right],
\end{eqnarray}
where $\epsilon\ll1$. It can be seen that that total Hamiltonian is unchanged,
\begin{eqnarray}
H_{A,B}'(t) &\equiv& A'(t)H'_\mathrm{drive} + B'(t)H'_\mathrm{prob} \nonumber \\
&=& A(t)H_\mathrm{drive} + B(t)H_\mathrm{prob},
\end{eqnarray}
but we have that
\begin{eqnarray}
B'(0) &=& \epsilon.
\end{eqnarray}
We define
\begin{eqnarray}
\Gamma'(t) &\equiv& \frac{A'(t)}{B'(t)} \\
\Gamma'(t) &=& \frac{\Gamma(t)}{\left[1+\Gamma(t)\frac{\epsilon}{A(0)}\right]}.
\end{eqnarray}
It can be immediately seen that $\Gamma'(t)$ is non-negative if $\Gamma(t)$ is non-negative, and so condition \ref{condition_nonnegative_Gamma}.~is satisfied.
Furthermore,
\begin{eqnarray}
\frac{\mathrm{d}\Gamma'(t)}{\mathrm{d}\Gamma(t)} &=& \frac{1}{\left[1+\Gamma(t)\frac{\epsilon}{A(0)}\right]^2}.
\end{eqnarray}
Thus, $\Gamma'(t)$ is monotonically-decreasing if $\Gamma(t)$ is is monotonically decreased, and so condition \ref{condition_monotonic_Gamma}.~is satisfied.

If we were to start the protocol in the state $\ket{\psi'_\mathrm{gs}}$, a ground-state of $H'_\mathrm{drive}$, condition \ref{condition_initial_ground_state}.~would be satisfied and the result would be proven.
However, the original protocol we are considering starts in the the state $\ket{\psi(0)}$, ground-state of $H_\mathrm{drive}$.
Applying first order perturbation theory in $\epsilon$ to $H'_\mathrm{drive}$, we find that $H'_\mathrm{drive}$ has a ground-state
\begin{eqnarray}
\ket{\psi'_\mathrm{gs}} &=& \ket{\psi(0)} + O\left(\frac{\epsilon}{A(0)\Delta}\right)\ket{\psi_\perp} \label{eq:psi_prime}
\end{eqnarray}
where $\ket{\psi_\perp}$ is a normalized state vector orthogonal to $\ket{\psi(0)}$ and $\Delta$ is the energy gap between the ground and first-excited manifolds of the actual driver Hamiltonian $H_\mathrm{drive}$.
Thus, assuming the driver Hamiltonian $H_\mathrm{drive}$ is not gapless (which is automatically true for all Hamiltonians on Hilbert spaces of finite dimension), the inequality in (\ref{eqn:quench_beats_guessing}) is satisfied in the limit as $\epsilon \rightarrow 0$.

\section{Lower bound on the average dynamic coefficient \label{app:dyn_bound}} %%%%%%%%%%%%%%%%%%%%%%%

\subsection{Bound on probabilities in a range based on second moment\label{app:moment_bound}}
Here, we prove a useful bound that will be applied in the following subsection. Assume that the distribution $p(x)$ has a finite second moment
\begin{equation}
\mu_2(p)=\int_{-\infty}^{\infty}\mathrm{d} x\,  p(x)(x-\mu_1(p))^2,
\end{equation}
where
\begin{equation}
\mu_1(p)=\int_{-\infty}^{\infty}\mathrm{d} x\,  p(x) x,
\end{equation}
is the first moment (mean). 
Let us choose some values $x_{\max} > x_{\min}$ such that $\mu_1(p)=\frac{1}{2}(x_{\max} + x_{\min})$.
The distribution $q(x)=\frac{1}{2}\delta(x_{\min}-\epsilon)+\frac{1}{2}\delta(x_{\max}+\epsilon)$ has the minimum possible second moment while having no support in the interval $\left[x_{\min}, x_{\max}\right]$, where $\delta$ is the Dirac delta distribution. 
In the limit $\epsilon \rightarrow 0$, the second moment of this distribution is $\mu_2(q)=\frac{1}{4}(x_{\max}-x_{\min})^2$. 
Thus, if $\mu_2(p)<\mu_2(q)$, then $p(x)$ must have some support within the range $\left[x_{\min},x_{\max}\right]$.
In particular, because second moment $\mu_2(p)$ can be lower bounded as 
\begin{eqnarray}
\mu_2 &=& \int_{-\infty}^{\infty}\mathrm{d} x\,  p(x)(x-\mu_1(p))^2 \nonumber \\
&\geq&  \int_{-\infty}^{x_{\min}}\mathrm{d} x\,  p(x)(x-\mu_1(p))^2 + \nonumber \\ 
& &\int_{x_{\max}}^{\infty}\mathrm{d} x\,  p(x)(x-\mu_1(p))^2 \nonumber \\
&\geq& \mu_2(q)\left(\int_{-\infty}^{x_{\min}}\mathrm{d} x\,  p(x) + \int_{x_{\max}}^{\infty}\mathrm{d} x\,  p(x)\right) \nonumber \\
&=&  \mu_2(q)\left(1 -  \int_{x_{\min}}^{x_{\max}}\mathrm{d} x\,  p(x)\right), \nonumber
\end{eqnarray}
the probability for $x$ to be in the interval $\left[x_{\min}, x_{\max}\right]$ can also be lower bounded as
\begin{eqnarray}
\int_{x_{\min}}^{x_{\max}}\mathrm{d} x\,  p(x) &\geq& 1 - \frac{\mu_2(p)}{\mu_2(q)} \nonumber \\
&=& 1- \frac{4\mu_2(p)}{(x_{\max} - x_{\min})^2}
\end{eqnarray}

\subsection{A simple lower bound}\label{app:dynbar_simplebound} %%%%%%%%%

Let $\zeta_{jk}=\frac{\left|\Delta_{jk}\right|}{2\left|\langle k|H_{\mathrm{drive}}|j\rangle \right|}$ and let $\eta_{jk}=\frac{\zeta_{jk}}{\Gamma}$. 
Furthermore, let $p_{\zeta}$ and $p_{\eta}$ be probability density functions that govern the distribution of the values $\zeta_{jk}$ and $\eta_{jk}$, respectively, over a set of problem instances. 
Let $\mu_{1}(p)$ and $\mu_{2}(p)$ refer
to the first and second moments, respectively, of a distribution governed by the probability density function $p$.

The dynamic coefficient is 
\begin{eqnarray}
\mathrm{Dyn}^{(jk)} & = & \frac{\eta_{jk}}{(1+\eta_{jk})^{2}},
\end{eqnarray}
so we will consider the function 
\begin{eqnarray}
f(x) & = & \frac{x}{(1+x)^{2}}
\end{eqnarray}
where $x>0$.

Let $x$ be distributed according to the probability density function $p_{\eta}$. We know that the expectation value $\langle f(x)\rangle_{x}$ is then 
\begin{eqnarray}
\langle f(x)\rangle_{x} &=& \intop_{0}^{\infty}\mathrm{d}xp_{\eta}(x)f(x)\nonumber\\
 & = & \intop_{x_{\min}}^{x_{\max}}\mathrm{d}xp_{\eta}(x)f(x)+\intop_{0}^{x_{\min}}\mathrm{d}xp_{\eta}(x)f(x)+\nonumber \\
 & & \intop_{x_{\max}}^{\infty}\mathrm{d}xp_{\eta}(x)f(x)\nonumber\\
 & = & P_{\eta}(x_{\min}<x<x_{\max})\langle f(x)\rangle_{x_{\min}}^{x_{\max}}+\nonumber \\
 & & P_{\eta}(x\geq x_{\min})\langle f(x)\rangle_{0}^{x_{\min}} + \nonumber\\ 
 & & P_{\eta}(x_{\max}\geq x)\langle f(x)\rangle_{x_{\max}}^{\infty}
\end{eqnarray}
where $x_{\max}>x_{\min}$, $P_{\eta}(\dots)$ is the probability of its argument being true if $\eta$ is distributed according to $p_{\eta}$, and $\langle f(x)\rangle_{a}^{b}$ is the expectation value of $f(x$) if $x$ is distributed according to a (renormalized) version of $p_{\eta}$ with all support on $x<a$ and $x>b$ removed. 
As $f(x)$ is positive for all $x>0$, we get the lower bound on $\langle f(x)\rangle_{x}$,
\begin{eqnarray}
 \langle f(x)\rangle_{x} & \geq & P_{\eta}(x_{\min}<x<x_{\max})\langle f(x)\rangle_{x_{\min}}^{x_{\max}} \nonumber\\
& > & P_{\eta}(x_{\min}<x<x_{\max})\min_{x_{\min}<x<x_{\max}}[f(x)].\nonumber
\end{eqnarray}
Since $f(x)$ is also convex, we know that 
\begin{eqnarray}
& \min_{x_{\min}<x<x_{\max}}[f(x)] = & \nonumber \\
& \min\left[f(x_{\min}),f(x_{\max})\right] &.
\end{eqnarray}

Now, let the interval $[x_{\min},x_{\max}]$ be of width $2c$ (for some $c>0$), and centred on the mean $\mu_{1}(p_{\eta})$ (thereby also constraining $c<\mu_{1}(p_{\eta})$). 
That is, $x_{\min}=\mu_{1}(p_{\eta})-c$ and $x_{\max}=\mu_{1}(p_{\eta})+c$, and we must find out which of $f(\mu_{1}(p_{\eta})-c)$ and $f(\mu_{1}(p_{\eta})+c)$ is smaller.
To do this, we consider under what conditions it is true that
\begin{eqnarray}
f(\mu_{1}(p_{\eta})-c) & < & f(\mu_{1}(p_{\eta})+c).\label{eqn:minus_version_smaller}
\end{eqnarray}
It can be shown that (\ref{eqn:minus_version_smaller}) is true when
\begin{eqnarray}
c^{2} & > & \mu_{1}^{2}(p_{\eta})-1
\end{eqnarray}
This inequality means that, when the mean $\mu_{1}(p_{\eta})$ is greater than $1$, the truth of the inequality in (\ref{eqn:minus_version_smaller}) depends on the value of $c$, but for $\mathrm{\ensuremath{\mu_{1}}(\ensuremath{p_{\eta}})}\leq1$, it is always true. 
Therefore, if we choose $\Gamma=\mu_{1}(p_{\zeta})$ (the mean of the distribution of $\zeta$ rather than $\eta$), then we have $\mu_{1}(p_{\eta})=1,$ which means the inequality $f(1-c)<f(1+c)$ is always true (where now $0<c<1$), and consequently 
\begin{eqnarray}
& \min_{\left(1-c\right)<x<\left(1+c\right)}[f(x)] = f(1-c) &\nonumber\\
& \langle f(x)\rangle_{1-c}^{1+c} > f(1-c) &\nonumber\\
& \langle f(x)\rangle_{x} >  P_{\eta}(1-c<x<1+c)f(1-c).&
\end{eqnarray}
Now, since $\zeta_{jk}\equiv\Gamma\eta_{jk}=\mu_{1}(p_{\zeta})\eta_{jk}$,
we have 
\begin{eqnarray}
P_{\eta}(1-c<x<1+c) = \nonumber\\
P_{\zeta}\left(\mu_{1}(p_{\zeta})\left(1-c\right)<x<\mu_{1}(p_{\zeta})\left(1+c\right)\right)
\end{eqnarray}
where $P_{\zeta}(\dots)$ is the probability of its argument being
true if $x$ is distributed according to $p_{\zeta}$.

Applying the result in subsection \ref{app:moment_bound}, we have 
\begin{eqnarray}
& P_{\zeta}\left(\mu_{1}(p_{\zeta})\left(1-c\right)<x<\mu_{1}(p_{\zeta})\left(1+c\right)\right) \geq & \nonumber \\ & 1-\frac{4\mu_{2}(p_{\zeta})}{\left(\left[\mu_{1}(p_{\zeta})\left(1+c\right)\right]-\left[\mu_{1}(p_{\zeta})\left(1-c\right)\right]\right)^{2}}
 = & \nonumber \\
 & 1-\frac{1}{c^{2}}\frac{\mu_{2}(p_{\zeta})}{\mu_{1}^{2}(p_{\zeta})}. &
\end{eqnarray}
Putting this all together gives 
\begin{eqnarray}
\max_\Gamma(\overline{\mathrm{Dyn}}) & \geq & f(1-c)\left[1-\frac{1}{c^{2}}\frac{\mu_{2}(p_{\zeta})}{\mu_{1}^{2}(p_{\zeta})}\right]\\
 & = & \frac{1-c}{(2-c)^{2}}\left(1-\frac{1}{c^{2}}\frac{\mu_{2}(p_{\zeta})}{\mu_{1}^{2}(p_{\zeta})}\right).
\end{eqnarray}
While this inequality gives a valid lower bound on $\overline{\mathrm{Dyn}}$,
the greatest lower bound can be written 
\begin{eqnarray}
& \max_\Gamma(\overline{\mathrm{Dyn}}) \geq & \nonumber \\
& \max_{0<c<1}\left[\frac{1-c}{(2-c)^{2}}\left(1-\frac{1}{c^{2}}\frac{\mu_{2}(p_{\zeta})}{\mu_{1}^{2}(p_{\zeta})}\right)\right],&
\end{eqnarray}
which can be found numerically for any given value of $\frac{\mu_{2}(p_{\zeta})}{\mu_{1}^{2}(p_{\zeta})}$ by optimizing over the parameter $c$.
This is plotted in Fig. \ref{fig:dyn_simplebound}.

\bibliography{QA_tools_for_rapid_quenches}  % bibtex entries in separate file for convenience

\end{document}